\documentclass[12pt]{article}

\usepackage[margin=0.8in]{geometry}
\usepackage{amsmath, amssymb, amsthm}
\usepackage{graphicx}
\usepackage{xcolor}
\usepackage{booktabs}
\usepackage{array}
\usepackage{multirow}
\usepackage{makecell}
\usepackage{tabularx}
\usepackage{hyperref}
\usepackage{natbib}
\usepackage{setspace}
\usepackage{caption}
\usepackage{comment}

\hypersetup{
    colorlinks=true,
    linkcolor=blue!60!black,
    citecolor=blue!60!black,
    urlcolor=blue!60!black
}

\newcommand{\avg}[1]{\langle #1 \rangle}
\newcommand{\beff}{\beta_{\rm eff}}
\newcommand{\geff}{\gamma_{\rm eff}}
\newcommand{\SIS}{\textsc{sis}}
\newcommand{\SIR}{\textsc{sir}}
\newcommand{\clSSI}{\frac{(n-1)\langle SS \rangle \langle SI \rangle}{\langle S \rangle}}
\newcommand{\clISI}{\frac{(n-1)\langle SI \rangle^2}{\langle S \rangle}}

\usepackage{rotating}   
\usepackage{mdframed}   

\newcommand{\av}[1]{\langle #1 \rangle}

\newcommand{\la}{\langle}
\newcommand{\ra}{\rangle}

\newcommand{\chg}[1]{\textcolor{red}{#1}}
\renewcommand{\chg}[1]{#1}

\newmdenv[
  backgroundcolor=blue!6,
  linecolor=blue!40,
  linewidth=1pt,
  innertopmargin=8pt,
  innerbottommargin=8pt,
  innerleftmargin=10pt,
  innerrightmargin=10pt,
  skipabove=10pt,
  skipbelow=6pt
]{recipebox}
 
\newmdenv[
  backgroundcolor=green!6,
  linecolor=green!45!black,
  linewidth=1pt,
  innertopmargin=6pt,
  innerbottommargin=6pt,
  innerleftmargin=10pt,
  innerrightmargin=10pt,
  skipabove=8pt,
  skipbelow=4pt
]{resultbox}

\newtheorem{remark}{Remark}
\newtheorem{prop}{Proposition}

\begin{document}

\begin{center}
    {\large \bfseries
    Fast-variable reduction to complex contagion unifies epidemic models
    }

    \vspace{1em}

    {
         Laurent H\'{e}bert-Dufresne$^{1,2,3}$,\quad
        P\'eter L.\ Simon$^{4}$,\quad
        Istv\'{a}n Z.\ Kiss$^{5,6,*}$
    }

    \vspace{0.6em}

    {\small
        $^{1}$Vermont Complex Systems Institute and Department of Computer Science, University of Vermont,
        Burlington, VT 05408, USA\\[2pt]
        $^{2}$Santa Fe Institute, Santa Fe, NM 87501, USA\\[2pt]
        $^{3}$Complexity Science Hub, 1030 Vienna, Austria\\[2pt]
        $^{4}$Institute of Mathematics, E\"{o}tv\"{o}s Lor\'{a}nd University,
        Budapest, Hungary\\[2pt] 
        $^{5}$ Network Science Institute, Northeastern University London, London, E1W 1LP, UK\\[2pt]
        $^{6}$ Department of Mathematics, Northeastern University, Boston, MA 02115, USA\\[2pt]
         $^{*}$correspondence: \href{mailto:istvan.kiss@nulondon.ac.uk}
        {istvan.kiss@nulondon.ac.uk}
    }
\end{center}

\vspace{1.2em}

\begin{abstract}
Across biological and social complex systems, models of epidemics or contagions incorporate complexity through network correlations, adaptive behavior, higher-order interactions, or co-circulating strains. 
Tracking these mechanisms lead to high-dimensional mathematical systems that resist analytical treatment.
We show that a broad class of models admits a systematic reduction via fast-variable elimination.
The central insight is that complexity can be captured by fast variables equilibrating much faster than the epidemic, leading to ``emergent complex contagion'' where the force of infection depends on prevalence.
The epidemic transition follows a universal equation and is continuous or discontinuous whether the force of infection decays or intensifies with prevalence.
Our results reveal diverse epidemic models share a common skeleton such that mechanisms that share a characteristic form can be macroscopically indistinguishable.
\end{abstract}

\section{Introduction}
\label{sec:intro}

Modern epidemics models have transformed our understanding of disease spread. 
They take the classic mass-action assumption where all agents interact identically and replace it with detailed transmission mechanisms and explicit contact or interaction structure~\citep{Kiss2017,PastorSatorras2015}.
Take the Susceptible-Infectious-Susceptible model, or \SIS\ for short, where infectious agents infect their susceptible neighbors at some rate $\tau$ and recover at some rate $\gamma$. 
In the classical pairwise \SIS\ model on networks, one can capture correlations between adjacent susceptible--infected pairs on a contact graph~\citep{Keeling1999,Eames2002} accounting for the statistical dependence that a network's structure induces between the states of neighboring agents.
Other models have since generalized the approach to capture broader correlations within triangles \cite{Barnard2019epidemic} or due to degree heterogeneity \cite{pastor2001epidemic, Eames2002, Kiss2023necessary}, capturing how local structure can substantially shift epidemic thresholds.
Subsequent extensions have embedded higher-order interactions via simplicial complexes \citep{Iacopini2019,Malizia2023}, allowed network topology to co-evolve with disease state in adaptive networks~\citep{Gross2006,Scarpino2016effect}, and explored the complex phenomenology of co-circulating pathogens~\citep{Hebert2015, Cai2015, Hebert2025one}.

Perhaps because of the richness of mechanisms considered, systematic comparison across model families remains largely qualitative.
Each model typically consists of a system of differential equations whose dimension grows with the order of correlations tracked, and whose epidemic threshold and bifurcation structure must be extracted by separate \emph{ad hoc} means.
Yet, the success of most mathematical models shows that local information (degree heterogeneity, clustering, node states) is often enough to capture the complex dynamics of epidemics on networks \citep{PastorSatorras2015, melnik2011unreasonable}.
This is in part because network effects and detailed transmission mechanisms tend to average out locally at a timescale much faster than the overall timescale of the epidemics across the entire system.

What follows is a unifying observation that motivates this paper. 
Despite their mechanistic diversity, the models above share a common mathematical fate: each can be written as a single equation in which the prevalence $i$ grows under an \emph{effective force of infection} $\beff(i)$ that is a function of prevalence $i$ measured by a reduced model,
\begin{equation}
   \dot{i} \;=\; i\,\bigl[\beff(i)-\geff(i)\bigr] \; ,
   \label{eq:beff_form_1}
\end{equation}
with $\geff(i)$ being the recovery rate, which is typically simply a constant rate $\gamma$.
This form is familiar from phenomenological models of nonlinear incidence~\citep{Liu1986, Liu1987dynamical, Hethcote1991some} and complex contagions~\citep{Guilbeault2018}, where the prevalence-dependent transmission rate is \emph{postulated}. 
Our contribution is to show that for a broad class of mechanistic network models the same form arises \emph{constructively} and that $\beff$ represents the equilibrated values of fast variables. 
In this sense, ``emergent complex contagions'' are a generic consequence of timescale separation rather than dependent on specific interactions \citep{Hebert2020, Stonge2021universal}, threshold effects \cite{Anttila2017mechanistic, Stonge2021universal}, or heterogeneity \citep{StOnge2024nonlinear, Hebert2026simpson, Tang2026emergent} as previously thought.

This reframing supplies a missing common structure to analyze epidemic models.
It makes the epidemic threshold a single universal condition at $i=0$, $\beff(0)=\gamma$, common to every model. 
And whether the epidemic emerges continuously or jumps discontinuously with bistability is decided simply by whether $\beff$ increases or decreases with prevalence at the threshold. 
The universal criterion for the critical point of Eq.~(\ref{eq:beff_form_1}) is detailed in Appendix \ref{sec:app_proposition}.

\begin{figure}[]
\centering
\includegraphics[width=0.65\textwidth]{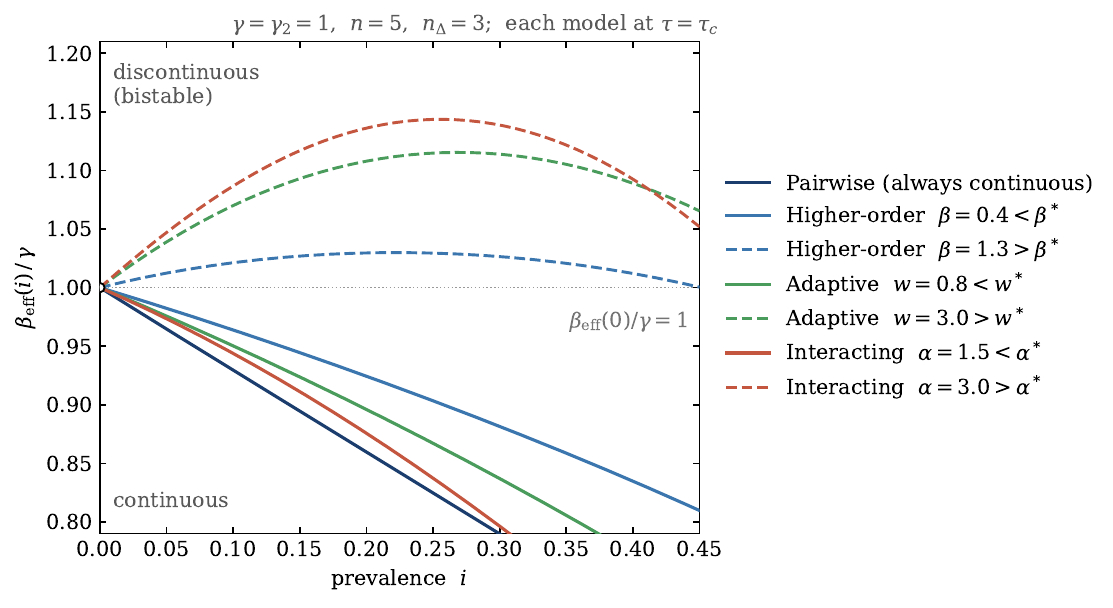}
\caption{\textbf{Complex-contagion skeleton on a single curve.} Effective force of infection $\beff(i)$, normalised by $\gamma$, plotted against
prevalence $i$ for four mechanistically distinct models, each evaluated \emph{at its
 own epidemic threshold} $\tau=\tau_c$.
 All curves pass through the universal threshold condition
$\beff(0)/\gamma=1$ of Eq.~\eqref{eq:universal_threshold}, found at the common point
$(0,1)$, as it is shared by every model
regardless of the underlying mechanisms. The sign of the slope at this point
classifies the transition through the criterion of
Eq.~\eqref{eq:bif_rule}: curves leaving $(0,1)$ \emph{downward}
($\partial_i\beff|_0<0$, solid) give a continuous transition, while curves
leaving \emph{upward} ($\partial_i\beff|_0>0$, dashed) signal the positive feedback
that produces a discontinuous transition with bistability.
}
\label{fig:beff_overlay}
\end{figure}

We present the fast-variable framework in general and apply it to five canonical
models: the classical pairwise \SIS\ model on regular networks,  the pairwise \SIS\ model with simplicial (three-body) interactions, an
adaptive-network model, a well-mixed model of two interacting \SIS\ contagions and the pairwise \SIR\
model.
Each is treated by the same four-step recipe. Figure~\ref{fig:beff_overlay}
previews the resulting collapse onto a
shared mathematical skeleton parameterised by $\beff(i)$; Table~\ref{tab:beff_skeleton}
summarises the corresponding thresholds and bifurcation conditions once the
necessary notation has been introduced (Section~\ref{sec:philosophy}), with full
derivations given in Appendix~\ref{sec:five_models}. Finally, we also show how the
recipe can be used numerically for models that defy analytical treatment.

\section{Unification through an effective force of infection}
\label{sec:philosophy}

The global prevalence $i$ is the slow, observable quantity. The correlations that make network models hard to study are fast.
We collect these fast quantities into a vector $\mathbf{p}$.
Every model in this paper then takes the generic form
\begin{equation}
   \dot{i} = i\,\bigl[ \beta(\mathbf{p},i) - \gamma(\mathbf{p},i)\bigr], \qquad \dot{\mathbf{p}} = \mathbf{F}(i,\mathbf{p}).
   \label{eq:generic_ip}
\end{equation}

For all canonical models in this paper, $\beta(\mathbf{p},i)$ is an average of a pair or joint state across infectious nodes. And for all but one model, $\gamma(\mathbf{p},i) = \gamma$, so we consider this case as the generic one.
The dimension of $\mathbf{p}$ is also immaterial to the principle: it is one for the classical pairwise, simplicial, interacting contagions, and SIR models, and two for the adaptive model, but in every case the construction is the same.

We see in Fig.~\ref{fig:fast_variable} that  $\mathbf{p}$ equilibrates much faster than the epidemic itself. Hence, we impose the quasi-static (adiabatic) condition $\dot{\mathbf{p}}=\mathbf{0}$ and solve the resulting algebraic system for $\mathbf{p}$ as a function of the slow variable $i$ alone, i.e. $\mathbf{p}(i)$. 
Introducing $\beff(i) = \beta(\mathbf{p}(i),i)$ and similarly $\geff(i)=\gamma(\mathbf{p}(i),i)$, we obtain the single self-contained differential equation, i.e., Eq.~\eqref{eq:beff_form_1}, hereafter the \emph{1D model}, which is the common skeleton onto which all five models map exactly around their epidemic threshold. 
Their mechanistic differences survive only as the functional form of their effective rates and the degree of the algebraic equation that defines them.

The epidemic threshold is determined by setting the growth rate of $i$ to zero at $i=0$, which provides the universal condition
\begin{equation}
   \beff(0) = \gamma .
   \label{eq:universal_threshold}
\end{equation}
The threshold is fixed entirely by the effective transmission rate, independently of the dimension of $\mathbf{p}$.
The nature of the transition, whether continuous or discontinuous with bistability, is a question about the reduced one-dimensional system of Eq.~\eqref{eq:generic_ip}. 
Writing it as $\dot x = f(x,\lambda)$ with bifurcation parameter $\lambda$, the disease-free state $x=0$ is an equilibrium for all $\lambda$, so $f$ factors as $f(x,\lambda)=x\,g(x,\lambda)$. 
A transcritical bifurcation occurs where $g(0,\lambda_c)=0$, and its direction is decided by a single ratio of derivatives of $g$ at that point. We state this as Proposition~\ref{prop:transcrit} in Appendix \ref{sec:app_proposition}; its content is that
\begin{equation}
   \frac{\partial_x g(0,\lambda_c)}{\partial_\lambda g(0,\lambda_c)} < 0
   \;\Rightarrow\; \text{continuous}, \qquad
   {} > 0 \;\Rightarrow\; \text{discontinuous} .
   \label{eq:bif_rule}
\end{equation}
Concretely, $\partial_\lambda g$ measures how fast the disease-free state loses stability as $\lambda$ is increased through the threshold, while $\partial_x g = \partial_i\beff|_{i=0}$ (where now $x=i$ and $g=\beff$) measures whether transmission intensifies or weakens with prevalence. 
A force of infection that decays with prevalence gives a continuous transition; one that grows with prevalence provides the positive feedback that produces a discontinuous transition and bistability. 

Most remarkably, a corollary of the fast-variable philosophy is that near threshold, every mechanism in this paper collapses through reduction onto the same one-dimensional skeleton.
Its transition type is set by the slope of the effective force of infection, and models that produce similar slopes are by construction indistinguishable. 
Thus, no macroscopic observation confined to the neighborhood of the epidemic threshold can identify the underlying mechanism. 

Figure~\ref{fig:beff_overlay} makes this collapse visible by evaluating every model at its own epidemic threshold $\tau_c$.
Every curve leaves $(0,1)$ with a definite sign of slope, and that sign alone decides whether the epidemic emerges gradually or jumps discontinuously into an endemic state, regardless of the complexity of the mechanisms or of the fast variables they produce. 
The figure also makes a subtler point concrete: three of the mechanisms shown (simplicial, adaptive, interacting contagions) are each plotted twice, once below and once above their own critical parameter ($\beta^{*}$, $w^{*}$, $\alpha^{*}$ respectively). 
The same mechanism can therefore sit on either side of the continuous/discontinuous divide depending on a single scalar, and two curves from \emph{different} mechanisms that happen to leave $(0,1)$ with the same slope are, at threshold, indistinguishable from one another. 
Table~\ref{tab:beff_skeleton} expands on this comparison by tabulating $\tau_c$, the leading-order behaviour of $\beff(i)$ near threshold, and the transition boundary for all five models once the pairwise \SIS\ worked example below has fixed the notation.

\section{Fast-variable reduction of canonical models\label{sec:reduction}}
We apply this recipe to five canonical models of epidemics on networks, each chosen to isolate a different source of complexity: pair correlations in the classical pairwise \SIS\ model, three-body interactions in simplicial contagion \cite{Malizia2025}, co-evolving topology in adaptive networks \cite{Gross2006}, strain interaction between co-circulating pathogens \cite{Hebert2015}, and a second slow variable in the pairwise \SIR\ model.
Treating all five within a single framework makes the generality of the method apparent.
One example requires us to introduce two fast variables, and another an effective force of recovery, but these generalizations are straightforward and do not change the general recipe.
In fact, the recipe extends naturally to further settings, such as contact tracing \cite{deMeijere2026trace} or the clustered pairwise SIS model \cite{Barnard2019epidemic}, which we do not develop in detail here. 

\begin{figure}[hp!]
\centering
\includegraphics[width=\textwidth]{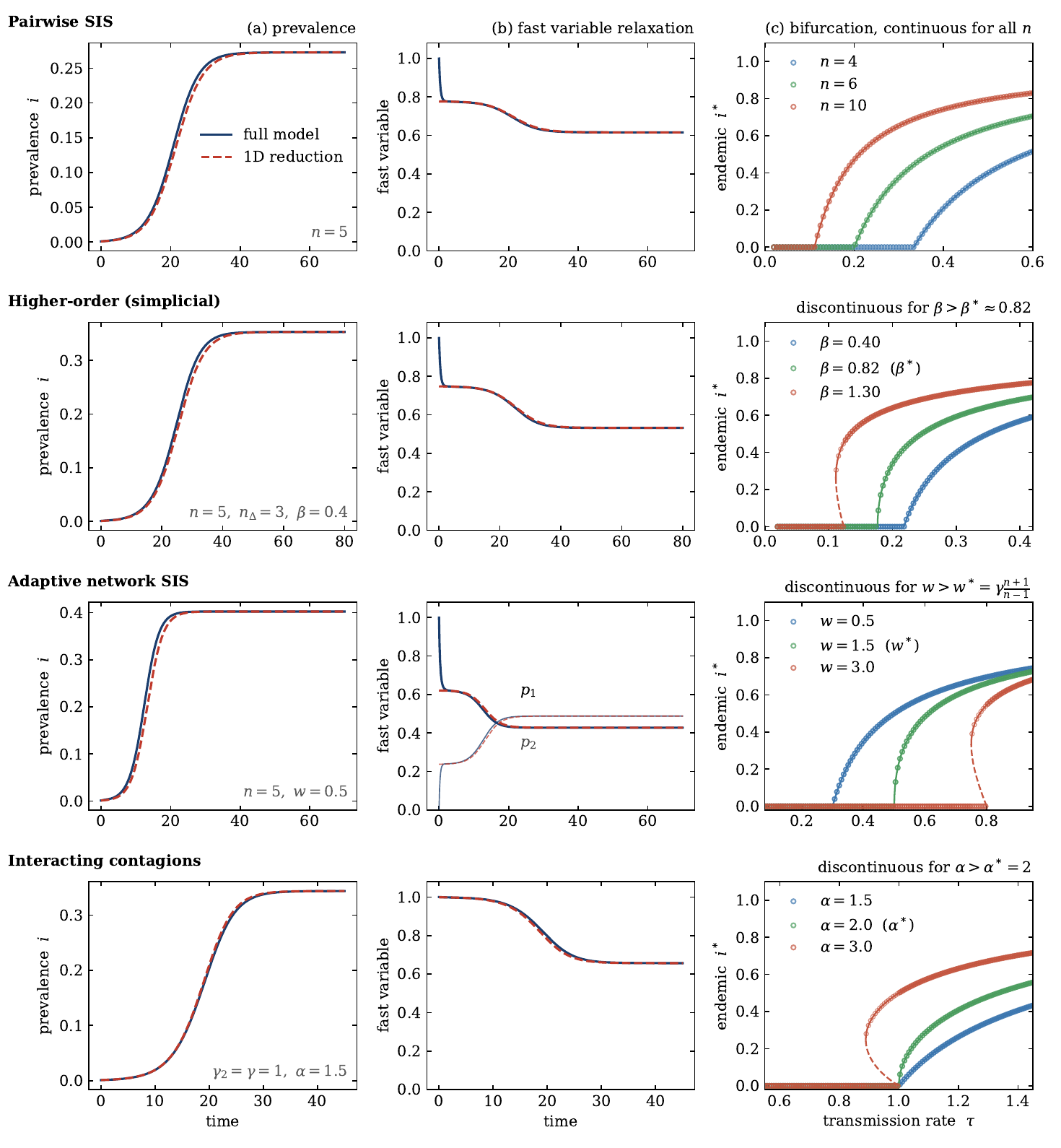}
\caption{\textbf{Fast-variable reduction across four endemic model families.}
Rows: pairwise \SIS, higher-order (simplicial) \SIS, adaptive network \SIS, and
interacting contagions.
Columns show (a) prevalence $i(t)$, (b) relaxation of the fast variable(s), (c) bifurcation diagrams of the full model (dots) with the analytic endemic branch from the reduced model (solid lines when stable, dashed when unstable).
In all panels, $\gamma=1$ while $\tau=1.3\,\tau_c$ in the time-series panels. }
\label{fig:fast_variable}
\end{figure}

We summarize the fast-variable reduction to the pairwise \SIS\ as our canonical example in the main text, and all full derivations can be found in Appendix \ref{sec:five_models}. The reduced models are visualized in Fig. \ref{fig:fast_variable}. 

\paragraph{Step 1 — Minimum networked SIS system.}
On a regular network of degree $n$, we track the expected fraction of susceptible nodes, $\avg{S}$, and infectious nodes, $\avg{I}$, as well as the expected fraction of directed pairs in a given joint-state $\avg{SS}$, $\avg{SI}$, or $\avg{II}$.
Using conservation laws
$\avg{S}+\avg{I}=1$, $\avg{SI}+\avg{II}=\avg{I}$ and the pair-closure
$\avg{ABC} = \avg{AB}\avg{BC}/\avg{B}$, the full system, see Eqs. \eqref{eq:PW_SIS_nReg_full_S}-\eqref{eq:PW_SIS_nReg_full_SS} in Appendix~\ref{sec:five_models},
reduces to two equations~\cite{Kiss2017,Eames2002}:
\begin{align*}
\dot{\avg{I}}   &= \tau n\avg{SI} -\gamma\avg{I},
\\
\dot{\avg{SI}}  &= \tau(n-2)\avg{SI}
- \frac{2\tau(n-1)\avg{SI}^2}{1-\avg{I}}
+\gamma\!\left(\avg{I}-2\avg{SI}\right).
\end{align*}
 
\paragraph{Step 2 — The $(i,p)$ system and the epidemic threshold.}
We set $i=\avg{I}$ and factor out $i$ in the prevalence equation to identify the fast variable $p=\avg{SI}/\avg{I}$. Differentiating $p$ using the minimum system leads to
\begin{align}
  \dot{i} &= \dot{\avg{I}} = i\,(\tau n p - \gamma),
  \label{eq:pw_SIS_ip_i_maintext}\\
  \dot{p} &= \frac{\mathrm{d}}{\mathrm{d}t}\left[\frac{\avg{SI}}{\avg{I}}\right] = \frac{\dot{\avg{SI}}}{\avg{I}} -p \frac{\dot{\avg{I}}}{\avg{I}} = \gamma (1-p) + \tau(n-2)p - \tau n p^{2} - 2\tau(n-1)\frac{p^{2}\,i}{1-i} \ .
  \label{eq:pw_SIS_ip_p_maintext}
\end{align}
At the disease-free state $(i,p)=(0,p_c)$, where $p_c$ is the solution of $0 = \gamma(1-p) + \tau(n-2)p - \tau n p^2$.
Plugging this into $\tau n p = \gamma$ leads to the following threshold condition $\tau_c=\gamma/(n-1)$ giving $p_c=(n-1)/n$. We note that this same threshold follows from the classical linear stability analysis of the full pairwise model.

\paragraph{Step 3 — 1D reduction.}
We use the quasi-static or adiabatic approximation by setting $\dot{p}=0$ in Eq.~\eqref{eq:pw_SIS_ip_p_maintext}. Defining $A(i)=\left[n+(n-2)i\right]/\left[n(1-i)\right]$ yields a quadratic equation for $p$ with positive root
\begin{equation*}
  p(i,\tau)=\frac{\tau(n-2)-\gamma
    +\sqrt{(\tau(n-2)-\gamma)^2+4\tau n\gamma\,A(i)}}
    {2\tau n\,A(i)}.
\end{equation*}
Equation \eqref{eq:pw_SIS_ip_i_maintext} thus becomes a one-dimensional system
\begin{equation*}
  \dot{i}
  =
  \beff^{\SIS}(i,\tau) \, i  -\gamma i, \,\,\, \text{where} \,\,\, \beff^{\SIS}(i,\tau)= \tau n p(i,\tau) \; .
\end{equation*} 
Since $A(i)$ is an increasing function of $i$, this result means that the pairwise \SIS~model maps to a sublinear complex contagion since $\beff^{\SIS}(i)$ decreases as prevalence $i$ increases (i.e., decreases faster than in the classic simple contagion). 
We can then use $\beff^{\SIS}(i)$ in our 1D model. Despite not capturing the extremely fast transient dynamics of $p$ because of the quasi-static approximation $\dot{p}=0$, the 1D model reproduces the full dynamics very well in Fig.~\ref{fig:fast_variable}.

\paragraph{Step 4 — Bifurcation.}
Now we apply Proposition \ref{prop:transcrit} with $g(i,\tau) = \tau n p(i,\tau)-\gamma$.
We can show that
\begin{equation*}
\frac{\partial_i g(0,\tau_c)}{\partial_{\tau} g(0,\tau_c) } = -\frac{ \gamma}{n} <0 \;.
\end{equation*}
Hence, Proposition \ref{prop:transcrit} implies that the bifurcation is of a continuous type and no discontinuous transition is possible in this model.

\paragraph{Generalizations.}
The same recipe can be used for the other four models shown in Appendix~\ref{sec:five_models}. All were chosen to introduce a different generalization that does not affect the recipe itself. The pairwise \SIR\ model can only reduce to a 2-dimensional system as both the fractions of susceptible and infectious are slow variables (with recovered nodes eliminated by conservation of population). The higher-order model with simplicial interactions in triangles leads to an implicit quartic equation for the fast variable. The adaptive network model requires two fast variables to capture all relevant correlations, but both are easily identified. And the interacting contagion model is parametrized as to require a prevalence-dependent effective recovery rate, which is shown not to affect our characterization of the epidemic threshold and its bifurcation.

Figure~\ref{fig:fast_variable} shows the results of this recipe applied to four different models (four rows). The first coulumn shows the prevalence $i(t)$ in the full system and in the reduced system at a transmission rate 30\% above the epidemic threshold to highlight the robustness of the reduction. The second column shows the time evolution of the fast variable $\mathbf{p}(t)$. The last column shows the different types of bifurcation curves obtained around the epidemic threshold $\beff(0,\tau)=\gamma$. 
Three things stand out. 
First, in every row the 1D reduction tracks the full model trajectory very well even away from $i=0$. 
Second, the fast variable relaxes to (or near) its quasi-static prediction $p^*(i)$ on a timescale visibly much shorter than the timescale over which $i(t)$ itself grows, confirming the timescale-separation assumption underlying our model reduction.
This holds even for the adaptive model, where two fast variables ($p_1,p_2$) must relax jointly.
Third, the bifurcation diagrams predicted analytically in Table~\ref{tab:beff_skeleton} 
are quantitatively correct: the analytic endemic branches $i^*$ from $\beff(i^*,\tau)=\gamma$ match the values obtained from the full model across the entire parameter range shown.
Where Fig.~\ref{fig:beff_overlay} shows that the skeleton $\beff(i)$ collapses several mechanisms onto a common skeleton near threshold, Fig.~\ref{fig:fast_variable} shows that the same reduction remains accurate for the full nonlinear dynamics away from threshold.

Table~\ref{tab:beff_skeleton} collects the outcome of this recipe applied to all five models. 
The full effective rate $\beff(i)$ is available in closed form only for some of them but in all cases it can be approximated as a power series.
Moreover, the two quantities that actually fix the qualitative behaviour are always closed form: the disease-free value $\beff(0)$, which sets the epidemic threshold, and the Proposition~\ref{prop:transcrit} ratio $\partial_i g/\partial_\tau g\big|_0$, whose sign classifies the transition type. 
A further point the table makes is that the transition type is not determined by the complexity of the fast-variable equation: the classical pairwise \SIS, adaptive, and interacting-contagion models all reduce to a quadratic in $p$, yet only the latter two admit a
discontinuous transition. 
The transition is fixed by the sign of the ratio, not by the degree of the characteristic equation.

\begin{table}[htbp]
\centering
\caption{\textbf{The five models on a common $\beff$ skeleton.}
Every model reduces to $\dot i = i[\beff(i)-\geff(i)]$ by quasi-static approximation of its fast
variable(s) $\mathbf p$. For each model we give $\beff$ as a function of the fast
variable (before the quasi-static approximation), the threshold $\tau_c$, the leading-order behavior of
$\beff(i)$ around $\tau_c$ and $i=0$, and the type of transition.
Equation references point to the derivations in Appendix~\ref{sec:five_models}.
}
\label{tab:beff_skeleton}
\renewcommand{\arraystretch}{1.6}
\setlength{\tabcolsep}{3pt}
\footnotesize
\begin{tabular}{%
    >{\raggedright\arraybackslash}p{2.3cm}
    >{\raggedright\arraybackslash}p{3.3cm}
    >{\raggedright\arraybackslash}p{2.2cm}
    >{\raggedright\arraybackslash}p{4.5cm}
    >{\raggedright\arraybackslash}p{2.8cm}
}
\toprule
\textbf{Model} &
\textbf{$\boldsymbol{\beff(p)}$ vs.\ fast variable $\boldsymbol{p}$} &
\textbf{Threshold $\boldsymbol{\tau_c}$} &
\textbf{$\boldsymbol{\beff(i)}$ near $\boldsymbol{\tau_c}$} &
\textbf{Type of transition} \\
\midrule
\textbf{Classical pairwise SIS} &
$\beff^{\SIS}=\tau n\,p$ &
$\tau_{c}=\frac{\gamma}{n-1}$ &
$\beff(i)=\gamma-\tfrac{2\gamma(n-1)^2}{n(2n-1)}\,i+O(i^2)$ &
always continuous \\
\midrule
\textbf{Higher-order pairwise SIS}\newline \scriptsize{simplicial $\beta$} &
$\beff^{\mathrm{HO}}=\tau n\,p$\newline
$\quad+\,\beta n_\triangle\frac{p^2(1-p)}{1-i}$ &
$\tau_c=\frac{\gamma}{n-1}-$\newline
$\beta n_\triangle\frac{n-2}{n^3}+O(\beta^2)$ &
$\beff(i)=\gamma\pm\tfrac{2\gamma\,\vert Q_n\vert}{P_n}\,i+O(i^2)$ \newline
from Eq.~\eqref{eq:dig-collapse} &
continuous for $\beta<\beta^{*}$\newline
discontinuous for $\beta>\beta^{*}=\frac{\gamma}{n_\triangle}h(n)$ \\
\midrule
\textbf{Adaptive network SIS}\newline \scriptsize{rewiring rate $w$} &
$\beff^{\mathrm{adap}}=\tau n\,p_1$ &
$\tau_c=\frac{\gamma+w}{n}$ &
$\beff(i)=\gamma+\tfrac{2\gamma\left[(n-1)w-(n+1)\gamma\right]}{\gamma(2n+1)+w}\,i+O(i^2)$, from Eq.~\eqref{eq:adap_dxg} &
continuous for $w<w^{*}$ \newline
discontinuous for $w>w^{*}=\gamma\frac{n+1}{n-1}$\\
\midrule
\textbf{Interacting contagions}\newline \scriptsize{synergy $\alpha$, $\gamma_2=\gamma$} &
$\beff^{\mathrm{inter}}=\tau\left[\alpha\!-\!(\alpha\!-\!1)p\right](1-i)$ &
$\tau_c=\gamma$ &
$\beff(i)=\gamma+\gamma\left(\alpha-2\right)i+O(i^2)$, \newline
from Eq.~\eqref{eq:ic_dxg} &
continuous for $\alpha<\alpha^{*}$ \newline
discontinuous for $\alpha>\alpha^{*}=2\gamma_2/\gamma$ \\
\midrule
\textbf{Classical pairwise SIR} &
$\beff^{\SIR}=\tau n\,p$ &
$\tau_{c}=\frac{\gamma}{n-2}$ &
$\beff(i)\approx\gamma-\tfrac{\gamma(n-1)}{n}\,i+O(i^2)$, \newline
from Eq.~\eqref{eq:sir_pofI} &
always continuous \\
\bottomrule
\end{tabular}
\end{table}


\section{Measuring the effective force of infection}
\label{sec:numerical}

Our analytical model reductions all rest on solving $\dot{\mathbf p}=\mathbf 0$ in closed form and obtaining $\mathbf p$ as a function of $i$, but that will not be possible for all models. 
This equation is already quartic for the simplicial model, and is likely to be unsolvable for more complicated ones.
This section shows that the effective force of infection $\beff(i)$ can instead be measured directly from a trajectory of the full model, without solving any algebraic equation for $\mathbf p$.

The procedure has two separable parts.
First, we fit the fast variable itself, $p(i)$, from the full model in steps 1-3 below.
Second, we push the fitted $\hat p(i)$ through the model-specific algebraic expression that defines $\beff$ in terms of $p$ and $i$, this leads to $\hat{\beta}_{\rm eff} (i)=\beta(\hat{p}(i),i)$ in step 4.

\paragraph{Numerical step 1 — Seed low.} Start at $i_0=10^{-6}$ for $t=0$. Seeding this low ensures that any transient behavior is over while $i$ is still small enough such that the low-prevalence region which determines the threshold and the bifurcation type remains rich in data.

\paragraph{Numerical step 2 — Discard the transient.} Retain the trajectory only after the fast variable(s) have equilibrated. In our code, that is when $|\dot p| < 5\,|\dot i|$ holds for $50$ consecutive samples. The criterion is crude but the results are insensitive to it because of the separation of timescales that the method exploits.

\paragraph{Numerical step 3 — Fit $p(i)$.} On the retained portion, regress the fast variable on the slow one with polynomials of order $K=0,1,2$,
        \begin{equation}
          \hat p(i)=\sum_{k=0}^{K} a_k\, i^{k}.
          \label{eq:polyfit}
        \end{equation}

\paragraph{Numerical step 4 — Close the model.} Substitute $\hat p(i)$ into the model-specific expression $\beff = \beta(p,i)$ (e.g.\ $\beff=\tau n\,\hat p(i)$ for pairwise \SIS) to obtain $\hat\beff(i)$, and solve the closed scalar equation $\dot i = i[\hat{\beta}_{\rm eff}(i)-\gamma]$.\\

\begin{figure}[hp]
\centering
\includegraphics[width=\textwidth]{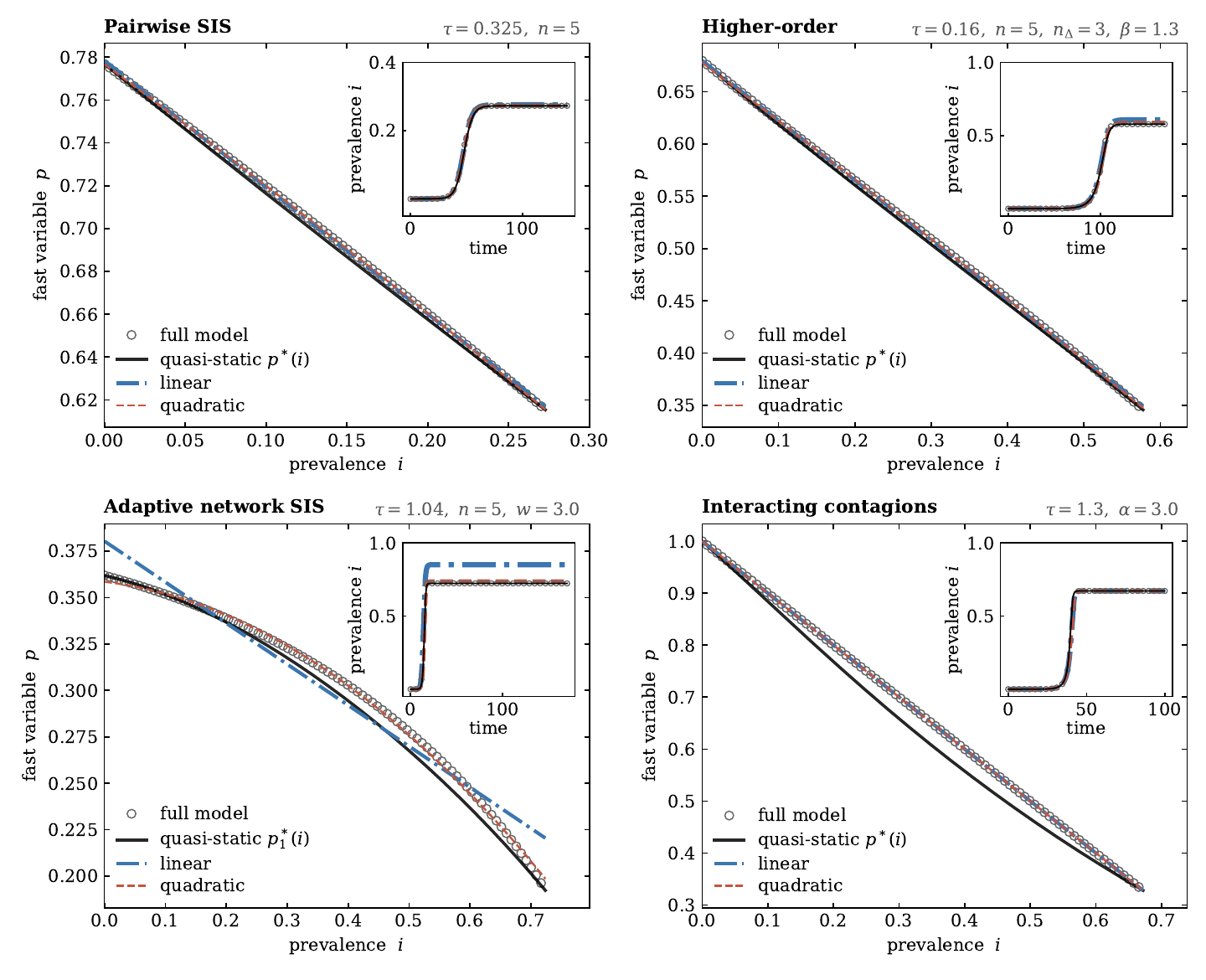}
\caption{\textbf{Measuring the fast variable.}
For each of the four endemic models, the fast variable $\mathbf{p}$ is plotted against the prevalence $i$ along a trajectory of the full model (circles), after the initial
transient has been discarded. Polynomial fits $\hat p(i)$ of degree $1$ (linear) and $2$ (quadratic) are shown, together with the analytic quasi-static prediction $p^{*}(i)$ (solid black). 
\emph{Insets:} prevalence $i(t)$ of the full model compared with the reduced models obtained by each fit.
Parameters are chosen to be in the discontinuous transition regime when possible.}
\label{fig:fitted_reduction}
\end{figure}

\paragraph{Interpretation.}
For example, for the pairwise \SIS~model, we have $\beff = \tau n\,p$ with $p=\avg{SI}/\avg{I}$. So, once $p(i)$ is known, $\beff(i)$ follows from a single multiplication by $\tau n$. We thus simulate the full \SIS\ system, discard the initial transient, and fit the retained $(i,p)$ pairs, this is the top-left panel of Fig.~\ref{fig:fitted_reduction}. A linear fit $\hat p(i)=a_0+a_1 i$ gives $\hat{\beta}_{\rm eff}(i)=\tau n(a_0+a_1 i)$, and substituting into $\dot i = i[\hat{\beta}_{\rm eff}(i)-\gamma]$ closes the reduced \SIS~model that reproduces the prevalence trajectory of the full model almost exactly (Fig.~\ref{fig:fitted_reduction}, inset). The full analytical derivation is not required. For the other three models the first step is unchanged; only the algebraic expression relating $p$ to $\beff$ in the second step differs, linear here, cubic in $p$ for the simplicial model, and so on.

We fit $p(i)$ rather than the time derivative $\dot i$ as a function of $i$ because $p$ is the quantity that equilibrates quickly and whose fitted slope carries the bifurcation information. In particular, tracking $a_0=p^{*}(0)$ can fix the threshold $\beff(0) = \gamma$. At this point, $a_1=\partial_i p^{*}|_0$ then fixes the bifurcation type through its effect on $\beff$ and the criterion of Section~\ref{sec:philosophy}.
A single linear regression close to $\tau_c$ can therefore recover both the threshold and the direction of the epidemic transition.

Figure~\ref{fig:fitted_reduction} fits all four endemic models and highlights two results.
First, the linear fits provide good approximations in most models.
The coefficients $a_0$ and $a_1$ shown in Table~\ref{tab:beff_skeleton} for the quasi-static approximation carry most of the information about the system, and not just at low $i$ as effective force of infection roughly maintains its shape for a large range of prevalence. 
This is in part why the one-dimensional reductions of the previous section perform as well as they do far from the bifurcation.
Second, the fits are always slightly above the analytic prediction $p^{*}(i)$, representing the small error introduced by the quasi-static approximation. The two always coincide at $i=0$ and at steady state, by definition, which is why the threshold is recovered exactly.

The results shown in Fig.~\ref{fig:fitted_reduction} rely on being able to run the full model. Thus, they are not a model reduction in the sense of Section~\ref{sec:reduction}. Still, it allows any complicated model whose fast variable resists closed-form elimination to be included in the same model-comparison framework.

Importantly, the procedure can also be applied on simulations or data without having to run the full mathematical model, as shown in Appendix~\ref{sec:simulations}. In which case the procedure does become a numerical model reduction that can facilite model selection.

\section{Conclusion}
\label{sec:discussion}

Equation~\eqref{eq:beff_form_1} re-frames all epidemic models discussed in the paper as complex contagions with prevalence-dependent rates.
Nonlinear incidence terms, often introduced phenomenologically \cite{Liu1986, Liu1987dynamical, Hethcote1991some} or in the context of complex contagions \cite{Guilbeault2018}, emerge here \emph{constructively} from dynamical correlations, network structure, or interactions.
The effective force of infection $\beff(i)$ thus provides a coarse-grained description. The details of microscopic mechanisms equilibrate quickly, and only the resulting transmission force of infection determines epidemic behavior.

We highlight this mapping and how it allows model analysis and comparison in Table~\ref{tab:beff_skeleton}.
For model analysis, our recipe yields analytical bifurcation insights for the adaptive, higher-order and interacting contagion models that were not uncovered by traditional techniques in their original presentation.
For model comparison, one critical consequence is a negative one. 
Near threshold the reduction collapses many models and mechanisms onto similar polynomial skeletons for the fast variable.
Models that share a skeleton are indistinguishable at their epidemic threshold.
It follows that no macroscopic observation confined to the neighborhood of the epidemic threshold, including the fact that a transition is discontinuous, can identify the mechanism that produced the epidemic transition. 

What does carry mechanistic information is the global shape of $\beff(i)$ away from the threshold.
This governs, for example, the width of the bistability window. 
These can differ between models even when their thresholds and transition types agree. 
The diagnostic value of the framework therefore lies not only in the bifurcation analysis but also beyond it as a model comparison tool.
Indeed, the effective forces of infection and their linear approximation shown in Table~\ref{tab:beff_skeleton} are linear in $\tau$ and contact density at $i=0$ for all models, but the slope $a_1$ of their dependency on $i$ contains much more information. Analysis of this coefficient across parameterizations is a very informative criterion for model selection. 

The recipe itself indicates where to go next. Its only requirement is a separation of timescales between global prevalence and the quantities that shape local transmission. Nothing in the construction limits the number or nature of the fast variables. 
Natural next targets include degree-heterogeneous networks, where the fast variable becomes a degree-resolved vector, and multilayer or bipartite contact structures, where inter-layer correlations contribute further fast variables. Our preliminary analysis of two-layer networks suggests that the reduction goes through when inter-layer coupling is weak, at the cost of a higher-degree characteristic equation. Models with vital dynamics or waning immunity may retain a second slow variable, as the pairwise \SIR\ model does. The limits of the approach are equally clear: the reduction is exact only at $i=0$, and it degrades wherever the separation of timescales fails. More broadly, the numerical procedure of Section~\ref{sec:numerical} requires only trajectories of the slow and fast variables. It therefore applies equally to stochastic simulations and, in principle, to agent-based models or to any system in which pair-level states can be observed. The same logic should carry over beyond epidemiology, to any spreading process, whether of opinions, behaviours or innovations, in which local structure equilibrates faster than the global state.

Complexity in epidemic models has usually been treated as something to be tracked: more correlations, more equations, more parameters. It can instead be absorbed. Once the fast processes have equilibrated, mechanisms leave behind a single prevalence-dependent force of infection, $\beff(i)$, that shapes the global dynamics. In this view, complex contagion is not a special modelling choice but the generic macroscopic signature of structured transmission. This changes the question facing model builders from \emph{which mechanism to include} to \emph{which effective force of infection is produced}, and whether data can tell the difference.

\section*{Author Contributions}

I.Z.K.\ and P.L.S.\ conceived the study and developed the fast-variable framework and classical pairwise analyses.
L.H.-D.\ contributed the adaptive network and interacting contagion analyses and the effective transmission rate framing.
All authors contributed to the manuscript and approved the final version.

\section*{Acknowledgements}

The authors thank Alessandro Vespignani and Giovanni Petri for comments on an early draft. I.Z.K.\ and L.H.-D.\ acknowledge discussions held at the CompleNet 2026 conference in Zaragoza (Spain) and NetSci 2026 conference in Boston (USA). This research has been implemented with the support from the National Research, Development and Innovation Office Hungary, financed under the ELTE TKP 2021-NKTA-62 funding scheme.

\section*{Competing Interests}

The authors declare no competing interests.

\section*{Data and Code Availability}

All analytical derivations are self-contained within the manuscript.
Computations were performed in Python. Codes are available at
\url{https://github.com/LaurentHebert/epi-fast-variables}.

\bibliographystyle{unsrt}

\section{Appendix}

\subsection{Transcritical bifurcation in 1D\label{sec:app_proposition}}

Consider the differential equation
\begin{equation}
    \dot x= f(x,\lambda),
    \label{eq:trans_crit_gen}
\end{equation}
where $x$ is a 1D state variable, $\lambda$ is a 1D parameter, and the trivial steady state $x=0$ exist for all values of $\lambda$, i.e. $f(0,\lambda)=0$ holds for any $\lambda$. In this case there is a function $g$ that $f$ can be factored as $f(x,\lambda)=xg(x,\lambda)$. The non-trivial steady state of the differential equation is given by the non-zero solution of $g(x,\lambda)=0$. We note that in the epidemic context, the trivial steady state is called disease-free, while the non-trivial is referred to as endemic steady state. Transcritical bifurcation occurs at the parameter value $\lambda=\lambda_c$, if $g(0,\lambda_c)=0$, that is the non-trivial steady state coincides with the trivial one. In the epidemic context, we are interested only in positive non-trivial steady states (i.e. $x>0$). In this sense two types of the transcritical bifurcation are used. The transcritical bifurcation is called forward, if the positive steady state exists after the bifurcation point, i.e. for $\lambda >\lambda_c$. (This means that the only non-negative steady state for $\lambda<\lambda_c$ is the trivial one $x=0$.) The transcritical bifurcation is called backward, if the positive steady state exists before the bifurcation point, i.e. for $\lambda <\lambda_c$. (This means that the trivial fixed point $x=0$ is the only non-negative steady state in a small neighbourhood of the origin for $\lambda>\lambda_c$.) Denoting the solution of $g(x,\lambda)=0$, i.e. the non-trivial branch of steady states, by $\lambda = \Lambda(x)$, the transcritical bifurcation is forward type, if $\Lambda'(0)>0$ and it is of backward type when $\Lambda'(0)<0$. This derivative can be easily computed by differentiating the implicit equation
\[
g(x,\Lambda(x))=0
\]
with respect to $x$ as
\[
\partial_x g(x,\Lambda(x)) + \partial_{\lambda} g(x,\Lambda(x)) \Lambda'(x) =0 .
\]
Hence substituting $x=0$ and using $\Lambda(0) = \lambda_c$, leads to
\[
\Lambda'(0) = - \frac{\partial_x g(0,\lambda_c)}{\partial_{\lambda} g(0,\lambda_c)} \ .
\]
The partial derivatives of $g$ can be related to those of $f$ by differentiating the relation $f(x,\lambda)=xg(x,\lambda)$ as follows. The first order partial derivatives are
\[
\partial_x f(x,\lambda)= g(x,\lambda) + x \partial_x g(x,\lambda), \qquad \partial_{\lambda} f(x,\lambda)= x\partial_{\lambda} g(x,\lambda) .
\]
Hence the condition $g(0,\lambda_c)=0$ is equivalent to $\partial_x f(0,\lambda_c)=0$.
The second order partial derivatives are
\[
\partial_{xx} f(x,\lambda)= 2\partial_x g(x,\lambda) + x \partial_{xx} g(x,\lambda), \qquad \partial_{\lambda x} f(x,\lambda)= \partial_{\lambda} g(x,\lambda) + x\partial_{\lambda x} g(x,\lambda) .
\]
Substituting  $x=0$ and $\lambda = \lambda_c$ yields
\[
\partial_{xx} f(0,\lambda_c)= 2\partial_x g(0,\lambda_c) , \qquad \partial_{\lambda x} f(0,\lambda_c)= \partial_{\lambda} g(0,\lambda_c)  ,
\]
enabling us to determine $\Lambda'(0)$ in terms of $f$. Thus we have the following
\begin{prop} \label{prop:transcrit}
    Consider differential equation \eqref{eq:trans_crit_gen} with 1-dimensional state variable $x$ and single parameter $\lambda$. Assume that $f(0,\lambda)=0$ holds for any $\lambda$, i.e. there is a trivial steady state independently of the parameter. If there is a parameter value $\lambda_c$ where $\partial_x f(0,\lambda_c)=0$, $\partial_{xx} f(0,\lambda_c)\neq 0$ and $\partial_{\lambda x} f(0,\lambda_c) \neq 0$, then transcritical bifurcation occurs at $\lambda = \lambda_c$. Moreover, we have the following.
    \begin{itemize}
        \item If $\frac{\partial_{xx} f(0,\lambda_c)}{\partial_{\lambda x} f(0,\lambda_c)} <0$, then the transcritical bifurcation is forward type.
        \item If $\frac{\partial_{xx} f(0,\lambda_c)}{\partial_{\lambda x} f(0,\lambda_c)} >0$, then the transcritical bifurcation is backward type.
    \end{itemize}
    The assumption $f(0,\lambda)=0$ implies that there is a function $g$ in such a way that $f$ can be expressed as $f(x,\lambda)=xg(x,\lambda)$. Hence the type of the bifurcation can be expressed in terms of $g$ as well as follows. 
    \begin{itemize}
        \item If $\frac{\partial_x g(0,\lambda_c)}{\partial_{\lambda} g(0,\lambda_c)} <0$, then the transcritical bifurcation is forward type.
        \item If $\frac{\partial_x g(0,\lambda_c)}{\partial_{\lambda} g(0,\lambda_c)} >0$, then the transcritical bifurcation is backward type.
    \end{itemize}
\end{prop}

\subsection{Fast-variable reduction of canonical models\label{sec:five_models}}
We choose pivotal formulation in the mathematical epidemiology of networks that together span a broad range of dynamics: the classical pairwise SIS model, the pairwise SIR model, higher-order (simplicial) contagion, adaptive networks, and multiple co-circulating strains. 
Treating all five within a single framework makes the generality of the method transparent, it is not tied to any one mechanism or contagion type, but applies uniformly across pair correlations, higher-order interactions, topological co-evolution, and strain interaction. 
One example requires us to introduce two  fast variables, and another an effective force of recovery, but these generalizations are straightforward and do not change the general recipe.
In fact, the recipe extends naturally to further settings, such as contact tracing \citep{deMeijere2026trace}, which we do not develop in detail here. 

For each model we begin with the most compact formulation, that is, the smallest set of equations that fully specifies the dynamics, followed by the derivation of the evolution equation(s) for the fast variable(s) together with the prevalence $i$ of the reduced system.
This new system, which at this stage remains equivalent to the original, is used to compute the epidemic threshold. 
Then, invoking the quasi-stationary approximation for the fast variable(s), we derive a one-dimensional equation for the prevalence $i$ governed by a model-specific effective transmission rate. 
Finally, the effective fast-variable system is used to analyse the bifurcation structure of each model.

\subsubsection{Classical pairwise SIS }
\label{sec:pairwise_main}

\paragraph{Step 0 — Full system.}
The full pairwise SIS equations are
\begin{align}
\dot{ \langle S \rangle} &= -\tau n\langle SI \rangle + \gamma \langle I \rangle, \label{eq:PW_SIS_nReg_full_S}\\
\dot{ \langle I \rangle} &= \tau n\langle SI \rangle - \gamma \langle I \rangle, \label{eq:PW_SIS_nReg_full_I}\\
\dot{ \langle SI \rangle} &=
 \tau (n-1)\langle SSI \rangle - \tau (n-1)\langle ISI \rangle -\tau \langle SI \rangle - \gamma \langle SI \rangle +\gamma \langle II \rangle, \label{eq:PW_SIS_nReg_full_SI}\\
 \dot{ \langle II \rangle} &=
 2 \tau (n-1)\langle ISI \rangle +2\tau \langle SI \rangle -2\gamma \langle II \rangle, \label{eq:PW_SIS_nReg_full_II}\\
 \dot{ \langle SS \rangle} &=-2
 \tau (n-1)\langle SSI \rangle + 2\gamma \langle SI \rangle,\label{eq:PW_SIS_nReg_full_SS}
\end{align}
with closure relations
\begin{equation}
    \langle SSI \rangle = \frac{\langle SS \rangle\langle SI \rangle}{\langle S \rangle} , \quad \langle ISI \rangle = \frac{\langle SI \rangle^2}{\langle S \rangle} \label{eq:PW_SIS_nReg_full_closures}
\end{equation}
see e.g.~\cite{Kiss2017,Eames2002}.
 
\paragraph{Step 1 — Minimum system.}
On a regular network of degree $n$, using conservation laws
$\avg{S}+\avg{I}=1$, $\avg{SI}+\avg{II}=\avg{I}$ and $\avg{SS}+\avg{SI}=\avg{S}$, the full 5-equation system,  Eqs.~\eqref{eq:PW_SIS_nReg_full_S}-\eqref{eq:PW_SIS_nReg_full_SS}, reduces to two equations
\begin{align}
\dot{\avg{I}}   &= \tau n\avg{SI} -\gamma\avg{I},
\label{eq:pw_SIS_min_I}\\
\dot{\avg{SI}}  &= \tau(n-2)\avg{SI}
- \frac{2\tau(n-1)\avg{SI}^2}{1-\avg{I}}
+\gamma\!\left(\avg{I}-2\avg{SI}\right).
  \label{eq:pw_SIS_min_SI}
\end{align}
 
\paragraph{Step 2 — The $(i,p)$ system and the epidemic threshold.}
Setting $i=\avg{I}$, $p=\avg{SI}/\avg{I}$, and then differentiating $p$ and using system 
\eqref{eq:pw_SIS_min_I}--\eqref{eq:pw_SIS_min_SI} lead to
\begin{align}
  \dot{i} &= i\,(\tau n p - \gamma),
  \label{eq:pw_SIS_ip_i}\\
  \dot{p} &= \gamma (1-p) + \tau(n-2)p - \tau n p^{2} - 2\tau(n-1)\frac{p^{2}\,i}{1-i}.
  \label{eq:pw_SIS_ip_p}
\end{align}
At the disease-free state $(i,p)=(0,p)$, $p$ is the solution of
\begin{equation}
0 = \gamma(1-p) + \tau(n-2)p - \tau n p^2 . 
\label{eq:pw_SIS_dfe}
\end{equation}
The Jacobian of system~\eqref{eq:pw_SIS_ip_i}--\eqref{eq:pw_SIS_ip_p} at the disease-free steady state $(0, p)$ is a lower triangular matrix. Its bigger eigenvalue is the number in the upper left corner: $\tau n p -\gamma$. Hence the threshold value of the parameters is determined by the equation
\begin{equation}
\tau n p -\gamma =0,
\label{eq:pw_SIS_threshold}
\end{equation}
where $p$ is determined by the quadratic equation \eqref{eq:pw_SIS_dfe}. The threshold condition for the parameters can be obtained by eliminating $p$ from system~\eqref{eq:pw_SIS_dfe}-\eqref{eq:pw_SIS_threshold}. This can be easily carried out by solving Eq.~\eqref{eq:pw_SIS_threshold} for $p$ as $p=\gamma/(\tau n)$ and plugging this into Eq.~\eqref{eq:pw_SIS_dfe}, leading to $p_c=(n-1)/n$. Substituting this value into Eq.~\eqref{eq:pw_SIS_threshold} the threshold condition is given as $\tau_c(n-1)=\gamma$. 
 
\paragraph{Step 3 — 1D reduction.}
Setting $\dot{p}=0$ in Eq.~\eqref{eq:pw_SIS_ip_p}, that is the quasi steady state approximation, and defining
\begin{equation}
A(i)=\frac{n+(n-2)i}{n(1-i)}
\label{eq:A_SIS}
\end{equation}
yields the quadratic $\tau n A(i)\,p^2-(\tau(n-2)-\gamma)p-\gamma=0$ with positive root
\begin{equation}
  p(i,\tau)=\frac{\tau(n-2)-\gamma
    +\sqrt{(\tau(n-2)-\gamma)^2+4\tau n\gamma\,A(i)}}
    {2\tau n\,A(i)}.
  \label{eq:pw_SIS_pofI}
\end{equation}
We note that it can be checked by simple algebra that this root is in the interval $[0,1]$. Hence, the one-dimensional equation in this is case is given by
\begin{equation}
  \dot{i}
  =
  \beff^{\SIS}(i,\tau) \, i  -\gamma i, \,\,\, \text{where} \,\,\, \beff^{\SIS}(i)= \tau n p(i,\tau).
  \label{eq:pw_SIS_1dEQ}
\end{equation} 
Evaluating $\beff^{\SIS}(0,\tau)-\gamma=0$ leads to the epidemic threshold, namely  $\tau(n-1)=\gamma$.
\paragraph{Step 4 — Bifurcation.}
Now we apply Proposition \ref{prop:transcrit} with $x=i$, $\lambda=\tau$ and $g(x,\lambda) = \tau n p(i,\tau)-\gamma$, where $p(i,\tau)$ is the solution of 
\begin{equation}
    \tau n A(i) p^2-(\tau(n-2)-\gamma)p-\gamma=0 .
    \label{eq:impl_p_SIS}
\end{equation}
%
According to Proposition \ref{prop:transcrit} we need the partial derivatives
\begin{equation}
\partial_x g(0,\lambda_c) = \tau_c n \partial_i p(0,\tau_c) \quad \mbox{ and } \quad \partial_{\lambda} g(0,\lambda_c) = n p(0,\tau_c) + \tau_c n \partial_{\tau} p(0,\tau_c) .
\label{eq:partial_g_p_SIS}
\end{equation}
Hence we differentiate the implicit Eq.~\eqref{eq:impl_p_SIS} with respect to $i$ and $\tau$ and then substitute $i=0$ and $\tau=\tau_c$. The derivative with respect to $i$ yields
\[
\tau n A'(i) p^2 + 2\tau n A(i) p \partial_i p  -(\tau(n-2)-\gamma) \partial_i p=0 .
\]
Substituting $i=0$ and $\tau=\tau_c=\gamma/(n-1)$ and using $p(0,\tau_c)=(n-1)/n$, $A(0)=1$ and $A'(0)=2(n-1)/n$ lead to
\begin{equation}
    \partial_i p (0,\tau_c) = -2 \frac{(n-1)^3}{n^2(2n-1)}.
    \label{eq:partial_i_p_SIS}
\end{equation}
 The derivative with respect to $\tau$ yields
\[
n A(i) p^2 + 2\tau n A(i) p \partial_{\tau} p -(n-2)p -(\tau(n-2)-\gamma) \partial_{\tau} p=0 .
\]
Substituting $i=0$ and $\tau=\tau_c$ and using $p(0,\tau_c)=(n-1)/n$, $A(0)=1$ lead to
\begin{equation}
    \partial_{\tau} p (0,\tau_c) = - \frac{(n-1)^2}{\gamma n (2n-1)} \; .
\label{eq:partial_tau_p_SIS}    
\end{equation}
Using Eq.~\eqref{eq:partial_g_p_SIS},  \eqref{eq:partial_i_p_SIS}, \eqref{eq:partial_tau_p_SIS} and $\tau_c=\gamma/(n-1)$ yields
\begin{equation}
\partial_x g(0,\lambda_c) = -2 \frac{(n-1)^3}{n^2(2n-1)} \tau_c n  = -2\gamma \frac{(n-1)^2}{n(2n-1)}
\label{eq:sis_dxg}
\end{equation}
and
\[
\partial_{\lambda} g(0,\lambda_c) = n-1 - \frac{(n-1)^2}{\gamma n (2n-1)} \tau_c n = n-1 - \frac{n-1}{2n-1} = 2 \frac{(n-1)^2}{2n-1}  . 
\]
Thus we have that
\[
\frac{\partial_x g(0,\lambda_c)}{\partial_{\lambda} g(0,\lambda_c) } = -\frac{ \gamma}{n} <0 ,
\]
hence Proposition \ref{prop:transcrit} implies that only a forward bifurcation occurs in the reduced 1-dimensional pairwise SIS model. 



\subsubsection{Higher-order pair-based SIS}
\label{sec:ho}
The original pair-based simplicial model of Malizia et~al.~\cite{Malizia2025} is written with pairwise rate $\beta$ and degree $k$,
simplicial rate $\beta_\triangle$ and triangle degree $k_\triangle$, and recovery
$\mu$. To keep notation consistent across this paper we relabel these as the
pairwise edge rate $\tau$, degree $n$, simplicial rate $\beta$, triangle degree
$n_\triangle$, and recovery $\gamma$, respectively. All equations below use this
convention.
\paragraph{Step 0 - Full system.} The higher-order pair-based SIS model is given in system (16) of Ref.~\cite{Malizia2025} as follows
\begin{align}
\dot{\langle S\rangle} =& \gamma \langle I\rangle - \tau \chg{n}\langle SI\rangle - \beta \chg{n_\triangle} \langle ISI^\triangle\rangle, \label{eq:PW_SIS_HO_full_S} \\[5pt]
\dot{\langle I\rangle} =& - \gamma \langle I\rangle + \tau \chg{n}\langle SI\rangle  + \beta \chg{n_\triangle}\langle ISI^\triangle \rangle, \label{eq:PW_SIS_HO_full_I}\\[5pt]
\dot{\langle SI\rangle} =& \gamma \langle II\rangle - \gamma \langle SI\rangle + \tau (\chg{n}-1)\langle SSI\rangle - \tau (\chg{n}-1)\langle ISI\rangle - \tau \langle SI\rangle \nonumber\\[5pt]
& + \beta \frac{\chg{n_\triangle}}{\chg{n}}(\chg{n}-2)\langle IIS^\triangle S\rangle - 2\beta \frac{\chg{n_\triangle}}{\chg{n}} \langle ISI^\triangle\rangle  - \beta\frac{\chg{n_\triangle}}{\chg{n}}(\chg{n}-2)\langle IIS^\triangle I\rangle, \label{eq:PW_SIS_HO_full_SI}\\[5pt]
\dot{\av{II}} =& - 2 \gamma \av{II} + 2 \tau (\chg{n}-1)\av{ISI} + 2 \tau\av{SI} + 4\beta \frac{\chg{n_\triangle}}{\chg{n}}\av{ISI^\triangle} + 2 \beta \frac{\chg{n_\triangle}}{\chg{n}}(\chg{n}-2) \av{IIS^\triangle I}\label{eq:PW_SIS_HO_full_II},\\[5pt]
\dot{\langle SS\rangle} =& 2\gamma\langle SI\rangle - 2\tau(\chg{n}-1)\langle SSI\rangle - 2\beta \frac{\chg{n_\triangle}}{\chg{n}}(\chg{n}-2) \langle IIS^\triangle S\rangle, \label{eq:PW_SIS_HO_full_SS}
\end{align}
with closures

\begin{align*} 
\av{ABC} &=\dfrac{\av{AB}\av{BC}}{\av{B}}, \quad \av{ABC^\triangle} = \dfrac{\av{AB}\av{BC}\av{AC}}{\av{A}\av{B}\av{C}}\\
\av{IIS^\triangle S} &= \dfrac{\av{SI}^2\av{II}\av{SS}}{\av{S}^2\av{I}^2},  \quad 
\av{IIS^\triangle I} = \dfrac{\av{SI}^3\av{II}}{\av{S}^2\av{I}^2} .
\end{align*}

We note that the above closure of quadruples assumes that the clustering coefficient $\Phi=0$. For the more general closure, depending on the value of $\Phi$, we refer to equation (15) in Ref.~\cite{Malizia2025}

\paragraph{Step 1 — Minimum system.}
It can be proved that the following conservation relations hold: $\langle S \rangle + \langle I \rangle=1 $ and $\la SS \ra +\la SI \ra=\la S \ra$, $\la SI \ra +\la II \ra=\la I \ra$. (Note that this proof exploits the special form of the closures.) Based on the conservations, we have 
 $\langle S \rangle =1- \langle I \rangle$, $\la II \ra =\la I \ra -\la SI \ra$ and $\la SS \ra =\la S \ra -\la SI \ra=1-\la I \ra -\la SI \ra$, reducing the full system of 5-equations to the following two equations.
\begin{align}
\dot{\langle I\rangle} =& - \gamma \langle I\rangle + \tau n\langle SI\rangle  + \beta n_\triangle \frac{\av{SI}^2(\av{I}-\av{SI})}{(1-\av{I})\av{I}^2},\label{eq:pw_SIS_HO_min_I}\\[5pt]
\dot{\langle SI\rangle} =& \gamma (\av{I}-\av{SI}) - \gamma \langle SI\rangle + \tau (n-1)\frac{(1-\av{I}-\av{SI})\av{SI}}{1-\av{I}} \nonumber\\[5pt]
& - \tau (n-1)\frac{
\av{SI}^2}{1-\av{I}} - \tau \langle SI\rangle 
+ \beta \frac{n_\triangle}{n}(n-2)\dfrac{\av{SI}^2(\av{I}-\av{SI})(1-\av{I}-\av{SI})}{(1-\av{I})^2\av{I}^2}  \nonumber \\[5pt]
&- 2\beta \frac{n_\triangle}{n} \frac{\av{SI}^2(\av{I}-\av{SI})}{(1-\av{I})\av{I}^2}- \beta\frac{n_\triangle}{n}(n-2)\dfrac{\av{SI}^3(\av{I}-\av{SI})}{(1-\av{I})^2\av{I}^2}. \label{eq:pw_SIS_HO_min_SI}
\end{align}
Note that this system is ill-defined at the disease-free steady state, where $\av{I}=0$ and $\av{SI}=0$. Hence in order to determine the Jacobian at the disease-free steady state -- that enables us to calculate the threshold value of the parameters -- we need to introduce new variables. Below we show that the fast variable can fill this role.

\paragraph{Step 2 — The $(i, p)$ system and the epidemic threshold.}
Setting $i = \langle I\rangle$ and $p = \langle SI\rangle/\langle I\rangle$, we have 
$$\dot  p =  \frac{\dot{\av{SI}}}{\av{I}} - p \frac{\dot{\av{I}}}{\av{I}}. $$ 
Using the minimal closed system~\eqref{eq:pw_SIS_HO_min_I}--\eqref{eq:pw_SIS_HO_min_SI}, leads to
\begin{align*}
\dot p
&= \gamma(1-p)
  +\tau(n-1)\frac{[1-i-2pi]\,p}{1-i}
  -\tau p - \tau n p^2
\notag\\
&\quad
  +\beta\frac{n_\triangle}{n}(n-2)
    \frac{p^2(1-p)[1-i-2pi]}{(1-i)^2} -2\beta \frac{n_\triangle}{n}\frac{p^2(1-p)}{1-i}
  - \beta n_\triangle\frac{p^3(1-p)}{1-i} \ .
\end{align*}

Simplifying this differential equation, the $(i, p)$  system takes the form
\begin{align}
\dot i &= i\left(
 \tau n p -\gamma 
+ \beta n_\triangle \frac{p^2(1-p)}{1-i}\right), \label{eq:pw_SIS_HO_ip_i}\\
\dot p
&= \gamma(1-p) +\tau(n-2)p - \tau n p^2- 2\tau(n-1)\frac{p^2i}{1-i} \notag\\
&\quad
  + \beta n_\triangle \frac{n-4-n p}{n}\frac{p^2(1-p)}{1-i} -2\beta n_\triangle \frac{n-2}{n} \frac{p^3(1-p)i}{(1-i)^2} \ .
\label{eq:pw_SIS_HO_ip_p}
\end{align}
It is important to note that this system is equivalent to system~\eqref{eq:pw_SIS_HO_min_I}--\eqref{eq:pw_SIS_HO_min_SI} but it is well-defined also for $i=0$. The disease-free steady state is $(i, p) = (0, p)$, the second coordinate of which is determined by the quartic equation
\begin{equation}
0 = \gamma(1-p) + \tau(n-2)p - \tau n p^2 + \beta n_\triangle \frac{n-4}{n}p^2(1-p) - \beta n_\triangle \,p^3(1-p). 
\label{eq:pw_SIS_HO_dfe}
\end{equation}
The Jacobian of system~\eqref{eq:pw_SIS_HO_ip_i}--\eqref{eq:pw_SIS_HO_ip_p} at the disease-free steady state $(0, p)$ is a lower triangular matrix. Its bigger eigenvalue is the number in the upper left corner: $\tau np -\gamma + \beta n_\triangle p^2(1-p)$. Hence the threshold value of the parameters is determined by the equation
\begin{equation}
\tau n p -\gamma + \beta n_\triangle p^2(1-p) =0,
\label{eq:pw_SIS_HO_threshold}
\end{equation}
where $p$ is determined by the quartic equation \eqref{eq:pw_SIS_HO_dfe}. The threshold condition for the parameters can be obtained by eliminating $p$ from system~\eqref{eq:pw_SIS_HO_dfe}-\eqref{eq:pw_SIS_HO_threshold}. This cannot be carried out as easily as for the pairwise SIS case, since the polynomial is higher order in $p$. Instead of a single equation that relates the threshold value of the parameters, one can derive the bifurcation curve in a parametric form with the parameter $p_c$. This can be achieved by exploiting that Eqs.~\eqref{eq:pw_SIS_HO_dfe} and \eqref{eq:pw_SIS_HO_threshold} are linear in the parameters, while non-linear in $p$. Substituting $\beta$ from Eq.~\eqref{eq:pw_SIS_HO_threshold} into Eq.~\eqref{eq:pw_SIS_HO_dfe} yields a linear equation for $\tau$ that can be solved as
\begin{equation}
\tau_c = \gamma \frac{2+n(p_c-1)}{np_c} \ .
\label{eq:ho_tau_crit}
\end{equation}

Plugging this expression into Eq.~\eqref{eq:pw_SIS_HO_threshold} leads to a linear equation in $\beta$ yielding
\begin{equation}
\beta_c = \gamma \frac{n(1-p_c)-1}{n_\triangle p_c^2(1-p_c)} \ .
\label{eq:ho_beta_crit}
\end{equation}
Since both $\tau$ and $\beta$ are non-negative, the value of $p_c$ can be varied in the interval $[1-2/n, 1-1/n]$. Note that at the right end point of the interval, $p_c=1-1/n$ we get back to the simple pairwise case when $\beta=0$ and $\tau (n-1)=\gamma$. Other values of $p_c$ yield the threshold value of $\tau$ for non-zero values of $\beta$.

Since this is an implicit relation between $\tau$ and $\beta$, it is useful to compute the series expansion of the above system around $\beta =0$. In order to do so, we introduce the inverse function $q$ that gives the value of $p_c$ from Eq.~\eqref{eq:ho_beta_crit} to a given value of $\beta_c$, that is $p_c = q(\beta_c)$. This can be substituted into Eq.~\eqref{eq:ho_tau_crit} to get the relation between $\tau_c$ and $\beta$ in the form
\[
\tau_c =\gamma + \gamma \frac{2-n}{nq(\beta)} \ .
\]
The linear approximation of this function around $\beta=0$ takes the form $\tau_c = \tau_c (0) + \tau_c'(0) \beta + O (\beta ^2)$, where
\begin{equation}
\tau_c(0) =\gamma + \gamma \frac{2-n}{nq(0)} \quad \mbox{ and } \quad \tau_c'(0) =- \gamma \frac{(2-n) q'(0)}{nq^2(0)} \ .
\label{eq:ho_tau_lin}
\end{equation}
Substituting $\beta =0$ into Eq.~\eqref{eq:ho_beta_crit} yields $q(0)=(n-1)/n$. Differentiating $\beta n_\triangle (q^2(\beta) - q^3(\beta))=\gamma n (1-q(\beta)) -\gamma$ according to $\beta$ and substituting $\beta =0$ leads to 
\[
q'(0)=-n_\triangle \frac{(n-1)^2}{\gamma n^4} \ .
\]
Substituting $q(0)$ and $q'(0)$ into Eq.~\eqref{eq:ho_tau_lin} yields the linear approximation
\[
\tau_c = \frac{\gamma}{n-1} - \beta n_\triangle \frac{n-2}{n^3} + O (\beta ^2) .
\]
We note that this reduces to the threshold $\tau_c(n-1)=\gamma$ of the SIS system when $\beta=0$.
\paragraph{Step 3 — 1D reduction.}
The quasi-static condition $\dot p = 0$ obtained from Eq.~\eqref{eq:pw_SIS_HO_ip_p} takes the form
\begin{equation}
  f_0(p) + \tau f_1(i,p) + \beta f_2(i,p) = 0,
  \label{eq:impl_p_SIS_HO}
\end{equation}
where 
\[
f_0(p) = \gamma(1-p) , \quad f_1(i,p) = (n-2)p - n p^2- 2(n-1)\frac{p^2i}{1-i} 
\]
and
\[
f_2(i,p) = n_\triangle \frac{n-4-np}{n}\frac{p^2(1-p)}{1-i} -2 n_\triangle \frac{n-2}{n} \frac{p^3(1-p)i}{(1-i)^2}.
\]
Equation \eqref{eq:impl_p_SIS_HO} is quartic in $p$ and admits no usable closed form for its roots. However, this quartic polynomial is positive when $p=0$ and it is negative, when $p=1$, hence we know that it has a root in the interval $[0,1]$. Let $p(i,\tau)$ denote a root in this interval determined implicitly by equation \eqref{eq:impl_p_SIS_HO}. Note that \eqref{eq:impl_p_SIS_HO} reduces to \eqref{eq:pw_SIS_HO_dfe} when $i=0$, hence $p(0,\tau_c)=p_c$.

Consequently the effective force of infection exists only implicitly: from the prevalence equation~\eqref{eq:pw_SIS_HO_ip_i},
\begin{equation}
  \beff^{\mathrm{HO}}(i,\tau) = \tau n \,p(i,\tau) + \beta  n_\triangle \frac{p(i,\tau)^{2}\bigl(1-p(i,\tau)\bigr)}{1-i}  ,
  \label{eq:ho_beff}
\end{equation}
leading to the 1D reduced equation
\begin{equation}
  \dot{i} =  \beff^{\mathrm{HO}}(i,\tau) \, i  -\gamma i .
  \label{eq:pw_SIS_HO_1dEQ}
\end{equation} 
Although $\beff^{\mathrm{HO}}(i,\tau)$ cannot be written down explicitly, its disease-free value — and hence the threshold and the transcritical bifurcation— is accessible. 

\paragraph{Step 4 — Bifurcation.}
Now we apply Proposition \ref{prop:transcrit},
\[
g(i,\tau) = \tau n p(i,\tau)-\gamma + \beta n_\triangle\frac{p(i,\tau)^{2}\bigl(1-p(i,\tau)\bigr)}{1-i},
\]
where $p(i,\tau)$ is the solution of Eq.~\eqref{eq:impl_p_SIS_HO} in $[0,1]$. According to Proposition \ref{prop:transcrit} we need the partial derivatives
\begin{equation}
\partial_i g(0,\lambda_c) = \tau_c n \partial_i p(0,\tau_c) + \beta_c n_\triangle  (2p_c-3p_c^2) \partial_i p(0,\tau_c) + \beta_c n_\triangle p_c^2(1-p_c) 
\label{eq:partial_g_p_i_SIS_HO}
\end{equation}
and
\begin{equation}
\partial_{\tau} g(0,\lambda_c) = n p_c + \tau_c n \partial_{\tau} p(0,\tau_c) + \beta_c n_\triangle (2p_c-3p_c^2) \partial_{\tau} p(0,\tau_c).
\label{eq:partial_g_p_tau_SIS_HO}
\end{equation}
Hence we differentiate the implicit  Eq.~\eqref{eq:impl_p_SIS_HO} with respect to $i$ and $\tau$ and then substitute $i=0$ and $\tau=\tau_c$. The derivative with respect to $i$ yields
\[
f_0'(p) \partial_i p  + \tau \partial_pf_1(i,p) \partial_i p + \tau \partial_i f_1(i,p)  + \beta \partial_p f_2(i,p) \partial_i p + \beta \partial_i f_2(i,p) = 0 
\]
leading to
\begin{equation}
\partial_i p = - \frac{\tau \partial_i f_1(i,p) + \beta \partial_i f_2(i,p) }{f_0'(p) + \tau \partial_pf_1(i,p) + \beta \partial_p f_2(i,p)} \ .
\label{eq:partial_i_p_SIS_HO_gen}
\end{equation}
Substituting $i=0$, $p=p_c$, $\beta = \beta_c$ and $\tau=\tau_c$ we have
\[
f_0'(p_c) = -\gamma , \quad \partial_p f_1(0,p_c) = n-2-2np_c, \quad \partial_p f_2(0,p_c) = \frac{n_\triangle}{n} \left[ 2p_c(n-4) -3p_c^2(2n-4) +4np_c^3 \right]
\]
and
\[
\partial_i f_1(0,p_c) = -2(n-1)p_c^2 , \quad \partial_i f_2(0,p_c) = \frac{n_\triangle}{n} \left( n-4 -3np_c +4p_c \right) p_c^2(1-p_c) .
\]
Substituting these partial derivatives into Eq.~\eqref{eq:partial_i_p_SIS_HO_gen} yields
\begin{equation}
    \partial_i p (0,\tau_c) = - \frac{A_1(p_c, \tau_c, \beta_c)}{C(p_c, \tau_c, \beta_c)} \ ,
    \label{eq:partial_i_p_SIS_HO}
\end{equation}
where
\[
A_1 (p_c, \tau_c, \beta_c)= -2\tau_cn (n-1)p_c^2 + \beta_cn_\triangle ( n-4 -3np_c +4p_c ) p_c^2(1-p_c)
\]
and
\[
C(p_c, \tau_c, \beta_c) = -\gamma n + \tau_c n (n-2-2np_c) + \beta_cn_\triangle \left[ 2p_c(n-4) -3p_c^2(2n-4) +4np_c^3 \right] .
\]
We note that $p_c$ is determined implicitly by $\tau_c$ and $\beta_c$ via system \eqref{eq:ho_tau_crit} -\eqref{eq:ho_beta_crit}.

The derivative with respect to $\tau$ yields
\[
f_0'(p) \partial_{\tau} p  +f_1(i,p) + \tau \partial_pf_1(i,p) \partial_{\tau} p + \beta \partial_p f_2(i,p) \partial_{\tau} p = 0 
\]
leading to
\begin{equation}
\partial_{\tau} p = - \frac{f_1(i,p) }{f_0'(p) + \tau \partial_pf_1(i,p) + \beta \partial_p f_2(i,p)} \; .
\end{equation}
Substituting $i=0$, $p=p_c$, $\beta = \beta_c$ and $\tau=\tau_c$ we have
\begin{equation}
    \partial_{\tau} p (0,\tau_c) = - \frac{A_2(p_c, \tau_c, \beta_c)}{C(p_c, \tau_c, \beta_c)} \; ,
\end{equation}
where
\[
A_2 (p_c, \tau_c, \beta_c)= n(n-2)p_c - n^2 p_c^2 .
\]
Using Eq.~\eqref{eq:partial_g_p_i_SIS_HO} yields
\[
\partial_x g(0,\lambda_c) = B_1(p_c, \tau_c, \beta_c) - D(p_c, \tau_c, \beta_c) \frac{A_1(p_c, \tau_c, \beta_c)}{C(p_c, \tau_c, \beta_c)},
\]
where
\[
B_1(p_c, \tau_c, \beta_c)= \beta_c n_\triangle p_c^2(1-p_c), \quad D(p_c, \tau_c, \beta_c)= \tau_cn + \beta_c n_\triangle (2p_c - 3 p_c^2) ,
\]
and Eq.~\eqref{eq:partial_g_p_tau_SIS_HO} yields
\[
\partial_{\lambda} g(0,\lambda_c) =  B_2(p_c, \tau_c, \beta_c) - D(p_c, \tau_c, \beta_c) \frac{A_2(p_c, \tau_c, \beta_c)}{C(p_c, \tau_c, \beta_c)} . 
\]
where
\[
B_2(p_c, \tau_c, \beta_c)= np_c .
\]
Thus we have that
\[
\frac{\partial_x g(0,\lambda_c)}{\partial_{\lambda} g(0,\lambda_c) } = \frac{ CB_1 -DA_1}{CB_2 -DA_2} \; .
\]



By Proposition~1 the transition changes character where
$\partial_i g(0,\lambda_c)/\partial_\tau g(0,\lambda_c)$ changes sign. We verify below
(Eq.~\eqref{eq:Rn-positive}) that $\partial_\tau g(0,\lambda_c)$ is strictly positive on
the threshold curve, so the boundary is the vanishing of the numerator alone,
\begin{equation}
  \partial_i g(0,\lambda_c) \;=\; \frac{C B_1 - D A_1}{C} \;=\; 0
  \qquad\Longleftrightarrow\qquad C B_1 = D A_1 .
  \label{eq:boundary-condition}
\end{equation}
This is one equation in the three unknowns $(p_c,\tau,\beta)$. The threshold curve
\eqref{eq:ho_tau_crit}--\eqref{eq:ho_beta_crit} supplies $\tau_c=\tau_c(p_c)$ and
$\beta_c=\beta_c(p_c)$; substituting them eliminates $\tau$ and $\beta$ and leaves a
single equation in $p_c$.

Two structural features are important. First, $\beta$ and
$n_\triangle$ enter the model only through the product $\beta n_\triangle$, and
Eq.~\eqref{eq:ho_beta_crit} fixes
$\beta_c n_\triangle\,p_c^{2}(1-p_c)=\gamma\left[n(1-p_c)-1\right]$, so
$n_\triangle$ disappears on substitution. Second, each of $A_1$, $B_1$, $C$, $D$ becomes
homogeneous of degree one in $\gamma$ once $\tau_c$ and $\beta_c$ are inserted, so
both sides of Eq.~\eqref{eq:boundary-condition} are $O(\gamma^{2})$ and $\gamma$ cancels.
The critical value $p_c$ is therefore a function of the pairwise degree alone.

Writing $\tau_c n = \gamma\,(np_c-n+2)/p_c$ and using the identity above, the four
quantities reduce to
\begin{align}
  B_1 &= \gamma\left[n(1-p_c)-1\right], \label{eq:B1-bar}\\[2pt]
  D   &= \gamma\,\frac{2np_c^{2}-3np_c+n+p_c}{p_c\,(1-p_c)}, \label{eq:D-bar}\\[2pt]
  A_1 &= \gamma\left[\,n(n-2)\,p_c^{2}-n(2n-5)\,p_c+(n-1)(n-4)\right], \label{eq:A1-bar}\\[2pt]
  C   &= -\,\gamma\,\frac{P_n(p_c)}{p_c\,(1-p_c)}, \label{eq:C-bar}
\end{align}
with
\begin{equation}
  P_n(p_c)=2n^{2}p_c^{3}-\left(5n^{2}-9n\right)p_c^{2}
              +\left(4n^{2}-15n+8\right)p_c-\left(n^{2}-6n+4\right).
  \label{eq:Pn}
\end{equation}
On the critical curve, $C$ can be rewritten so that every term carries the common denominator $p_c(1-p_c)$. This was done using Eq.~\eqref{eq:ho_tau_crit} to replace $\tau_c$ by adding an extra $(1-p_c)$ factor, and with  Eq.~\eqref{eq:ho_beta_crit} to replace $p_c\beta_cn_\triangle$ in the final term.

Combining Eqs.~\eqref{eq:B1-bar}--\eqref{eq:C-bar} gives
\begin{equation}
  C B_1 - D A_1 \;=\; \frac{2\gamma^{2}}{p_c\,(1-p_c)}\;Q_n(p_c),
  \qquad\text{hence}\qquad
  \partial_i g(0,\lambda_c) \;=\; -\,\frac{2\gamma\,Q_n(p_c)}{P_n(p_c)},
  \label{eq:dig-collapse}
\end{equation}
where
\begin{equation}
  Q_n(p_c)\;=\;2n^{2}p_c^{4}-\bigl(3n^{2}-n\bigr)p_c^{3}
  +2n\,p_c^{2}+\bigl(2n^{2}-5n+2\bigr)p_c-\bigl(n^{2}-3n+2\bigr)\; .
  \label{eq:ho_boundary_poly}
\end{equation}
Since $P_n(p_c)>0$ throughout
$p_c\in[1-2/n,\,1-1/n]$, the sign of $\partial_i g$ is opposite to that of $Q_n$: the
transition is \emph{forward} where $Q_n(p_c)>0$ and \emph{backward} where
$Q_n(p_c)<0$.

It remains to confirm that the denominator discarded in Eq.~\eqref{eq:boundary-condition} is
harmless. Substituting the threshold curve into $\partial_\tau g=(CB_2-DA_2)/C$ gives
\begin{equation}
  \partial_\tau g(0,\lambda_c)=\frac{2n\,p_c\,R_n(p_c)}{P_n(p_c)},
  \qquad
  R_n(p_c)=2np_c^{2}-(4n-3)p_c+2(n-1)=2n(1-p_c)^{2}+3p_c-2,
  \label{eq:dtaug}
\end{equation}
so that the full Proposition~1 ratio reads
\begin{equation}
  \frac{\partial_i g(0,\lambda_c)}{\partial_\tau g(0,\lambda_c)}
  = -\,\frac{\gamma\,Q_n(p_c)}{n\,p_c\,R_n(p_c)} .
  \label{eq:full-ratio}
\end{equation}
The factor $P_n$ cancels between numerator and denominator and therefore plays no role in
the classification. The quadratic $R_n$ has discriminant
\begin{equation}
  \operatorname{disc}R_n=(4n-3)^{2}-16n(n-1)=9-8n<0 \qquad (n\ge 9/8),
  \label{eq:Rn-positive}
\end{equation}
such that the denominator never crosses zero with a real $p_c$ on the admissible interval $p_c\in[1-2/n,\,1-1/n]$.
The denominator of Eq.~\eqref{eq:full-ratio} is therefore strictly positive and the sign of the ratio is fixed by $Q_n$ alone. 
As a consistency check, setting $\beta=0$ and $p_c=(n-1)/n$ in Eq.~\eqref{eq:dtaug} returns
$\partial_\tau g=2(n-1)^{2}/(2n-1)$, in agreement with our independent calculation for the
classical pairwise model.


A root of $Q_n$ in the admissible interval exists for every $n>2$, as the endpoint values
are available in closed form:
\begin{equation}
  Q_n\!\left(1-\tfrac1n\right)=\frac{(n-1)^{2}}{n^{2}}>0,
  \qquad
  Q_n\!\left(1-\tfrac2n\right)=-\,\frac{(n-2)\left(n^{2}-6n+12\right)}{n^{2}}<0,
  \label{eq:Qn-endpoints}
\end{equation}
the second because $n^{2}-6n+12$ has negative discriminant and is therefore positive for
all $n$. The positive value at $p_c=1-1/n$ is the consistency check with the classical
pairwise model: $\beta=0$ at that endpoint and the transition must be forward. Because $\beta_c$ increases monotonically as $p_c$
decreases from $1-1/n$, the sign change at $p_c^{\,*}$ is exactly the statement that
the transition is forward for $\beta<\beta^{*}$ and backward for $\beta>\beta^{*}$.

Solving $Q_n(p_c)=0$ numerically gives
$p_c^{\,*}=0.62076,\;0.72907,\;0.78977,\;0.88733,\;0.94721$ for
$n=4,5,6,10,20$, and substituting into Eq.~(29) returns the values of $h(n)$ quoted in
the main text.

Remarkably, $Q_n$ contains neither $n_\triangle$ nor $\gamma$: the critical value
$p_c^{*}$ is a function of the pairwise degree \emph{alone}. Substituting
$p_c^{*}$ back into Eq.~\eqref{eq:ho_beta_crit} therefore gives the simplicial rate at
which the transition changes character in closed form,
\begin{equation}
  \beta^{*}=\frac{\gamma}{n_\triangle}\,h(n),
  \qquad
  h(n)=\frac{n\bigl(1-p_c^{*}\bigr)-1}{p_c^{*2}\bigl(1-p_c^{*}\bigr)}\;
  \label{eq:ho_betastar}
\end{equation}
with $p_c^{*}$ the root of $Q_n$ in $[1-2/n,\,1-1/n]$. The transition is forward for
$\beta<\beta^{*}$ and backward (bistable) for $\beta>\beta^{*}$. Thus $\beta^{*}$ is
\emph{exactly} proportional to $\gamma/n_\triangle$, and all of the dependence on the
pairwise degree is carried by the single scalar $h(n)$, which decreases monotonically
($h(4)=3.538$, $h(5)=2.463$, $h(6)=1.993$, $h(10)=1.428$, $h(20)=1.177$): denser
pairwise contact structure makes bistability easier to reach. For $n=5$, $n_\triangle=3$
and $\gamma=1$ this gives $\beta^{*}=0.8209$, in agreement with the numerically located
boundary used in Fig.~\ref{fig:fast_variable}.

\subsubsection{Adaptive network SIS}
\label{sec:adaptive_corrected}

Following Gross, D'Lima \& Blasius~\citep{Gross2006}, we consider a classic SIS model with a pairwise description but adding a mechanism allowing susceptible nodes to rewire away from infected neighbours at rate $w$.
Following the original formulation, the minimum system has three variables $(I, \avg{II}, \avg{SS})$, which give rise to two fast variables $\mathbf{p} = [p_1, p_2]$.

\paragraph{Step 0 --- Full system.}

The adaptive network pairwise SIS model takes the form
\begin{align}
  {\dot{\avg{I}}}  &= {\tau n\,\avg{SI} - \gamma\avg{I},}  \label{eq:adaptC_I_App}\\[2pt]
  {\dot{\avg{SS}}} &= {2(\gamma+w)\avg{SI} - 2\tau n\,\frac{\avg{SI}\avg{SS}}{1-\avg{I}},}
    \label{eq:adaptC_SS_App}\\[2pt]
  {\dot{\avg{II}}} &= {2\tau\avg{SI}  + 2\tau n\,\frac{\avg{SI}^{2}}{1-\avg{I}} - 2\gamma\avg{II},}
    \label{eq:adaptC_II_App} \\[2pt]
    {\dot{\avg{SI}}} &= {\gamma\avg{II} -( \tau+\gamma + w)\avg{SI} + \tau n\,\frac{\avg{SI}(\avg{SS}-\avg{SI})}{1-\avg{I}} .}      \label{eq:adaptC_SI_App}
\end{align}
Note that our formulation of classical SIS pairwise model in Eqs.~\eqref{eq:PW_SIS_nReg_full_S}-\eqref{eq:PW_SIS_nReg_full_SS} corresponds to a regular network with degree $n$, while this model assumes a Poisson degree distribution. This is why the triple closure terms contain a factor $n$ here and $n-1$ there.

\paragraph{Step 1 --- Minimum system.}
Because rewiring conserves links, the arc-conservation law holds,
\begin{equation}
  \avg{SS} + \avg{II} + 2\avg{SI} = 1,
  \label{eq:adaptC_cons}
\end{equation}
together with $\avg{S}=1-\avg{I}$. With the pair closure
$\avg{ABC}=\avg{AB}\avg{BC}/\avg{B}$, the dynamics reduce to the three-variable
system
\begin{align}
  \dot{\avg{I}}  &= \tau n\,\avg{SI} - \gamma\avg{I},
    \label{eq:adaptC_I}\\[2pt]
  \dot{\avg{II}} &= 2\tau\avg{SI}
    + 2\tau n\,\frac{\avg{SI}^{2}}{1-\avg{I}}
    - 2\gamma\avg{II},
    \label{eq:adaptC_II} \\[2pt]
    \dot{\avg{SI}} &= \gamma\avg{II} -( \tau+\gamma + w)\avg{SI} + \tau n\,\frac{\avg{SI}(\avg{SS}-\avg{SI})}{1-\avg{I}}.
    \label{eq:adaptC_SI}
\end{align}
The rate $w$ enters only Eq.~\eqref{eq:adaptC_SI}, as the outflow that rewiring sends into the $\avg{SS}$ pool; $w=0$ recovers classical pairwise SIS. Unlike the static case, the stub identity
$\avg{SI}+\avg{II}=\avg{I}$ is broken by rewiring, so we have an extra degree of freedom and the minimum system is genuinely three-dimensional.

\paragraph{Step 2 --- The $(i,p_1,p_2)$ system and the epidemic threshold.}
Set $i=\avg{I}$, $p_1=\avg{SI}/\avg{I}$, $p_2=\avg{II}/\avg{I}$, so that
$\avg{SI}=p_1 i$, $\avg{II}=p_2 i$ and, by \eqref{eq:adaptC_cons},
$\avg{SS}=1-(2p_1+p_2)i$. Differentiating each ratio and using
Eqs.~\eqref{eq:adaptC_I}--\eqref{eq:adaptC_SI} gives the exact system
\begin{align}
  \dot i   &= i\,(\tau n\,p_1 - \gamma),
    \label{eq:adaptC_idot}\\[4pt]
  \dot p_1 &= \gamma p_2 + (\tau (n -1)-w)p_1 -\tau n p_1^2+ \tau n p_1(1-3p_1-p_2) \frac{i}{1-i} ,
    \label{eq:adaptC_p1dot}\\[4pt]
  \dot p_2 &= -\gamma p_2+2\tau p_1 -\tau n p_1p_2 + 2\tau n p_1^{2} \frac{i}{1-i}  .
    \label{eq:adaptC_p2dot}
\end{align}

At $i=0$ the steady state conditions $\dot p_1=\dot p_2=0$ become
\begin{align}
  0 &= \gamma p_2 - \tau n\,p_1^{2} + (\tau n-\tau-w)p_1,
    \label{eq:adaptC_dfe1}\\
  0 &= 2\tau p_1 - \gamma p_2 - \tau n\,p_1 p_2.
    \label{eq:adaptC_dfe2}
\end{align}
Equation~\eqref{eq:adaptC_dfe2} expresses the infected--infected correlation as a function of $p_1$,
\begin{equation}
  p_2 = \frac{2\tau p_1}{\gamma + \tau n\,p_1},
  \label{eq:adaptC_p2star}
\end{equation}
and substituting into Eq.~\eqref{eq:adaptC_dfe1} (discarding $p_1=0$) gives 
\[
\frac{2\tau \gamma}{\gamma + \tau n\,p_1} - \tau n\,p_1 + \tau n-\tau-w =0,
\]
leading to
\[
2\tau \gamma - \tau n\,p_1 (\gamma + \tau n\,p_1) + (\tau n-\tau-w)(\gamma + \tau n\,p_1)=0 ,
\]
which is the following quadratic in $p_1$
\begin{equation}
(\tau (n+1) -w) \gamma + \tau n\,p_1 (\tau n-\tau-w-\gamma) -\tau^2 n^2 p_1^2=0.
\label{eq:adaptC_p1_quadr}
\end{equation}
Imposing the threshold condition $\tau n\,p_1=\gamma$ from Eq.~\eqref{eq:adaptC_idot}, i.e.\
$p_1=\gamma/(\tau n)$, the above equation collapses it to $\gamma(\tau n-\gamma-w)=0$, hence the
invasion threshold is
\begin{equation}
  \tau_c n = \gamma + w.
  \label{eq:adaptC_threshold}
\end{equation}
Rewiring raises the threshold linearly in $w$, recovering the qualitative
result of Gross et~al.~\citep{Gross2006}.

\paragraph{Step 3 -- 1D reduction.}
The quasi-steady state assumption now takes the form $\dot p_1=0$, $\dot p_2=0$. Without assuming $i=0$ this leads to
\begin{align}
  0 &= \gamma p_2 + (\tau (n -1)-w)p_1 -\tau n p_1^2+ \tau n p_1(1-3p_1-p_2) \frac{i}{1-i} ,
    \label{eq:adaptC_p1_ss}\\
  0 &= -\gamma p_2+2\tau p_1 -\tau n p_1p_2 + 2\tau n p_1^{2} \frac{i}{1-i}  .
    \label{eq:adaptC_p2_ss}
\end{align}
Multiplying Eq.~\eqref{eq:adaptC_p2_ss} by $i/(1-i)$ and subtracting the two equations we can eliminate the term $p_1p_2$ and we can easily express $p_2$ in terms of $p_1$ as
\[
\gamma p_2 = \tau n p_1^2 \frac{1+i}{1-i} - p_1 ((\tau-w)(1-i) +\tau (n-2)) .
\]
This can be substituted into Eq.~\eqref{eq:adaptC_p2_ss} yielding a quadratic equation for $p_1$ (after division by $p_1$) with the following coefficients,

\begin{equation}
A=(n\tau)^2\frac{1+i}{1-i},\qquad
B=\tau n\bigl[\gamma+w-\tau(n-1)+i(\tau-w)\bigr],\qquad
C=\gamma\bigl[w-\tau(n+1)+i(\tau-w)\bigr] .\label{eq:ABC}
\end{equation}
Note that this quadratic condition reduces to Eq.~\eqref{eq:adaptC_p1_quadr} when $i=0$. Since $A>0$ and $C<0$ near threshold, the quadratic has exactly one non-negative root, moreover $A+B+C>0$ near the threshold yielding that the non-negative root is in $[0,1]$, so we keep
\begin{equation}
 p_1(i,\tau) = \frac{-B + \sqrt{B^2 - 4AC}}{2A} \in [0,1] \; .
\label{eq:adaptC_pofI}
\end{equation}
Hence, the one-dimensional quasi-steady state reduction is given by
\begin{equation}
  \dot{i}  =  \beff^{\mathrm{adap}}(i,\tau) \, i  -\gamma i, \,\,\, \text{where} \,\,\, \beff^{\mathrm{adap}}(i,\tau)= \tau n p_1(i,\tau).
  \label{eq:adaptC_1dEQ}
\end{equation} 
We note that evaluating $\beff^{\mathrm{adap}}(0,\tau)-\gamma=0$ leads to the epidemic threshold, namely  $\tau_c n=\gamma +w$.

\paragraph{Step 4 -- Bifurcation.} Now we determine the type of the transcritical bifurcation that occurs at this threshold value. We apply Proposition \ref{prop:transcrit} with $x=i$, $\lambda=\tau$ and $g(x,\lambda) = \tau n p_1(i,\tau)-\gamma$, where $p_1(i,\tau)$ is given by Eq.~\eqref{eq:adaptC_pofI}. That is, $p_1(i,\tau)$ is the solution of 
\begin{equation} 
A(i,\tau) p_1^2(i,\tau) + B(i,\tau) p_1(i,\tau) + C(i,\tau)=0 ,
\label{eq:adapt_p1_i_tau}
\end{equation}
where $A, B$ and $C$ are given in Eq.~\eqref{eq:ABC}.

We have shown above that the bifurcation occurs at $\lambda_c=\tau_c=(\gamma +w) / n$ and $p_1(0,\tau_c)=\gamma/(\gamma +w)$. According to Proposition \ref{prop:transcrit} we need the partial derivatives
\begin{equation*}
\partial_x g(0,\lambda_c) = \tau_c n \partial_i p_1(0,\tau_c) \quad \mbox{ and } \quad \partial_{\lambda} g(0,\lambda_c) = n p_1(0,\tau_c) + \tau_c n \partial_{\tau} p_1(0,\tau_c) .
\end{equation*}
We differentiate the implicit Eq.~\eqref{eq:adapt_p1_i_tau} with respect to $i$ and $\tau$ and then substitute $i=0$ and $\tau=\tau_c$. The derivative with respect to $i$ yields
\begin{equation*} 
\partial_i A(i,\tau) p_1^2(i,\tau) +  2A(i,\tau) p_1(i,\tau) \partial_i p_1(i,\tau)  + \partial_i B(i,\tau) p_1(i,\tau) + B(i,\tau) \partial_i p_1(i,\tau) + \partial_i C(i,\tau)=0 ,
\end{equation*}
hence
$$
\partial_i p (i,\tau) = - \frac{\partial_i A(i,\tau) p_1^2(i,\tau) + \partial_i B(i,\tau) p_1(i,\tau) + \partial_i C(i,\tau)}{2 A(i,\tau) p_1(i,\tau) + B(i,\tau) } \ .
$$
Substituting $i=0$ and $\tau=\tau_c$ and using
$$
A(0,\tau_c)= (\gamma + w)^2, \quad \partial_i A(0,\tau_c)= 2(\gamma + w)^2, \quad B(0,\tau_c)= \frac{1}{n}(\gamma + w)^2,
$$
$$
\partial_i B(0,\tau_c)= \frac{\gamma + w}{n}(\gamma - (n-1)w), \quad C(0,\tau_c)= -\frac{\gamma}{n}((n+1)\gamma + w), \quad \partial_i  C(0,\tau_c)= \frac{\gamma}{n}(\gamma -(n-1) w),
$$
leads to
\begin{equation}
\partial_i p (0,\tau_c) = - \frac{2\gamma}{\gamma +w} \frac{(n+1)\gamma - (n-1) w}{(2n+1)\gamma + w}.
\label{eq:partial_i_p_adapt}
\end{equation}
The derivative with respect to $\tau$ yields
\begin{equation*}
\partial_\tau A(i,\tau) p_1^2(i,\tau) +  2A(i,\tau) p_1(i,\tau) \partial_\tau p_1(i,\tau)  + \partial_\tau B(i,\tau) p_1(i,\tau) + B(i,\tau) \partial_\tau p_1(i,\tau) + \partial_\tau C(i,\tau)=0 ,
\end{equation*}
hence
$$
\partial_\tau p (i,\tau) = - \frac{\partial_\tau A(i,\tau) p_1^2(i,\tau) + \partial_\tau B(i,\tau) p_1(i,\tau) + \partial_\tau C(i,\tau)}{2 A(i,\tau) p_1(i,\tau) + B(i,\tau) } \ .
$$
Substituting $i=0$ and $\tau=\tau_c$ and using
$$
A(0,\tau_c)= (\gamma + w)^2, \quad \partial_\tau A(0,\tau_c)= 2n(\gamma + w), \quad B(0,\tau_c)= \frac{1}{n}(\gamma + w)^2,
$$
$$
\partial_\tau B(0,\tau_c)= (2-n)(\gamma + w), \quad C(0,\tau_c)= -\frac{\gamma}{n}((n+1)\gamma + w), \quad \partial_\tau  C(0,\tau_c)= -\gamma (n+1),
$$
leads to
\begin{equation}
\partial_{\tau} p (0,\tau_c) = - \frac{n \gamma}{(\gamma +w)^2} \frac{(2n-1)w -\gamma}{(2n+1)\gamma + w}.
\label{eq:partial_tau_p_adapt}
\end{equation}
Hence
the following formulas are derived for the partial derivatives of $g$.
\begin{equation}
\partial_x g(0,\lambda_c) = -2 \gamma  \frac{(n+1)\gamma-(n-1)w}{(\gamma +w)((2n+1)\gamma +w)}(\gamma +w)  = 2 \gamma  \frac{(n-1)w - (n+1)\gamma}{(2n+1)\gamma +w}
\label{eq:adap_dxg}
\end{equation}
and
\[
\partial_{\lambda} g(0,\lambda_c) = \frac{n\gamma}{\gamma +w} + \gamma n \frac{(2n-1)w -\gamma}{(\gamma +w)^2((2n+1)\gamma +w)} (\gamma +w) =    2\gamma \frac{n^2 }{(2n+1)\gamma +w}. 
\]
Thus we have that
\begin{equation}
\frac{\partial_x g(0,\lambda_c)}{\partial_{\lambda} g(0,\lambda_c) } = \frac{(n-1)w -(n+1)\gamma}{n^2} \ .
\label{eq:adap_partial_x_lam_g}
\end{equation}
Finally, Proposition \ref{prop:transcrit} implies that the transcritical bifurcation in the reduced 1D equation \eqref{eq:adaptC_1dEQ} is forward (continuous) when $w(n-1)<\gamma(n+1)$ and backward (discontinuous, with bistability) when $w(n-1)>\gamma(n+1)$. The two regimes are separated by
\begin{equation*}
w^\ast=\gamma\,\frac{n+1}{n-1}.
\end{equation*} 

The endemic branch of the reduced differential equation \eqref{eq:adaptC_1dEQ} can also be determined analytically as follows. The endemic steady state $i^*$ of Eq.~\eqref{eq:adaptC_1dEQ} satisfies $\beff^{\mathrm{adap}}(i^*,\tau)=\gamma$, which is equivalent to $\tau n p_1(i^*,\tau)=\gamma$. This can be substituted into the quadratic equation for $p_1$, leading to the following relation between $i^*$ and $\tau$.
\[
\gamma^2 \frac{1+i^*}{1-i^*} + \gamma (\gamma +w -\tau (n-1) + i^*(\tau-w)) +\gamma (w -\tau (n+1) +i^*(\tau-w)) =0 .
\]
This is linear in $\tau$, hence $\tau$ can be given as a function of the endemic prevalence $i^*$ as
\begin{equation}
  \tau(i^*) = \frac{\gamma + w\,(1-i^*)^{2}}{(1-i^*)(n-i^*)} \ .
  \label{eq:adaptC_branch}
\end{equation}
Its $i\to0$ limit reproduces the threshold of Eq.~\eqref{eq:adaptC_threshold}, while the sign of $\tau'(0)$ classifies the transition:
\begin{equation}
  \tau'(0) = \frac{\gamma(n+1)-w(n-1)}{n^{2}} \ .
  \label{eq:adaptC_slope}
\end{equation}
The branch turns forward (continuous, supercritical) when $\tau'(0)>0$ and backward (discontinuous, with a saddle-node fold and bistability) when $\tau'(0)<0$. Note that Eq.~\eqref{eq:adaptC_slope} is identical to the condition given in Eq.~\eqref{eq:adap_partial_x_lam_g}, keeping in mind that according to Proposition \ref{prop:transcrit} we have $\tau'(0) = - \partial_x g/ \partial_{\lambda} g$. 

\subsubsection{Interacting contagions}
\label{sec:interacting}
Two \SIS\ diseases with co-infection enhancement $\alpha$ are considered, following
H\'{e}bert-Dufresne et al.~\citep{Hebert2020}. Nodes that are co-infected with both diseases can produce a higher transmission rate (e.g. superspreading events) and different recovery rates (e.g. compromised immune system). The fast variable is the
singly-infected fraction, i.e., the fraction of nodes infected with a given disease but not the other, $p=\avg{SI}/\avg{I}$.

\paragraph{Step 1 — Minimum system.}
We consider a well-mixed model of two interacting \SIS\ contagions. Each individual
occupies one of four joint states $\avg{SS}$, $\avg{IS}$, $\avg{SI}$, $\avg{II}$,
where the first index denotes the status with respect to disease~1 and the second
with respect to disease~2, subject to the normalisation
\begin{equation}
    \avg{SS} + \avg{IS} + \avg{SI} + \avg{II} = 1.
\end{equation}
Being infectious with one disease multiplies the rate at which an individual
transmits the other by $\alpha$ (e.g.\ through a diminished immune response), and
co-infected individuals clear \emph{each} of their two infections at rate
$\gamma_2$, whereas singly-infected individuals recover at rate $\gamma$. The case
$\gamma_2<\gamma$ represents a prolonged co-infection period; $\gamma_2=\gamma$
returns the standard model.

We set $\tau_1=\tau_2=\tau$ and take initial conditions symmetric between the two
diseases, so that $\avg{IS}=\avg{SI}$ for all time and the two prevalences coincide,
$\avg{I}\equiv\avg{IS}+\avg{II}=\avg{SI}+\avg{II}$. Writing
$\avg{SS}=1-2\avg{SI}-\avg{II}$ and collecting the force of infection into
\begin{equation}
  \Phi \;=\; \tau\bigl(\avg{SI}+\alpha\avg{II}\bigr),
  \label{eq:ic_Phi}
\end{equation}
the model for disease 1 reduces to the two-dimensional system
\begin{align}
    \dot{\avg{II}} &= 2\avg{SI}\,\Phi \;-\; 2\gamma_2\avg{II},
    \label{eq:II_sym}\\[4pt]
    \dot{\avg{SI}} &= \bigl(1-2\avg{SI}-\avg{II}\bigr)\Phi
                   \;+\; \gamma_2\avg{II} \;-\; \gamma\avg{SI}
                   \;-\; \avg{SI}\,\Phi .
    \label{eq:IS_sym}
\end{align}

\paragraph{Step 2 — The $(i,p)$ system and the epidemic threshold.}
Set $i=\avg{I}$ and $p=\avg{SI}/\avg{I}$, so that $\avg{SI}=p\,i$ and, by conservation of infections with disease~1, $\avg{II}=(1-p)\,i$. 
The force of infection of Eq.~\eqref{eq:ic_Phi} then factorises as $\Phi=i\tau\bigl[\alpha-(\alpha-1)p\bigr]$ while the two distinct recovery rates combine into an emergent, prevalence-dependent
\emph{effective recovery rate}
\begin{equation}
  \gamma_{\mathrm{eff}}^{\mathrm{inter}}(i) \;=\; \gamma\,p + \gamma_2\,(1-p),
  \label{eq:ic_geff}
\end{equation}
the population-weighted clearance rate of the infected class. Adding
Eqs.~\eqref{eq:II_sym}--\eqref{eq:IS_sym} and using $\avg{SS}+\avg{SI}=1-i$ gives the
prevalence equation:
\begin{equation}
  \dot i \;=\; i\,\bigl[\beff^{\mathrm{inter}}(i) - \gamma_{\mathrm{eff}}^{\mathrm{inter}}(i)\bigr]
  \label{eq:ic-idot}
\end{equation}
with the effective force of infection
\begin{equation}
  \beff^{\mathrm{inter}}(i) \;=\; \tau\bigl[\alpha-(\alpha-1)p\bigr](1-i).
  \label{eq:ic_beff}
\end{equation}
The factor $(1-i)$ stands out as different from other models seen so far, but appears because we are dealing with a well-mixed population such that $(1-i)$ captures the susceptible neighbors of infected nodes. 
From there, the quotient rule $\dot p=\dot{\avg{SI}}/i - p\,\dot i/i$ then yields the exact fast
equation
\begin{equation}
  \dot p \;=\; \dfrac{\beff^{\mathrm{inter}}(i)}{1-i}\bigl[\,(1-p)-(1+p)\,i\,\bigr]
        \;+\; (1-p)\bigl[\gamma_2(1+p)-\gamma p\bigr].
  \label{eq:ic-pdot}
\end{equation}
Both reduce to the familiar symmetric expressions when $\gamma_2=\gamma$, where
$\gamma_{\mathrm{eff}}^{\mathrm{inter}}\equiv\gamma$ and the bracket in Eq.~\eqref{eq:ic-pdot} collapses to
$\gamma(1-p)$.

This is the first model in our catalogue in which the \emph{recovery} side of the
prevalence equation is itself driven by a fast variable. The generic
form of Eq.~\eqref{eq:generic_ip} is unchanged, but the universal threshold condition
$\beff(0)=\gamma$ of Section~\ref{sec:philosophy} appears here in its natural
generalisation,
\begin{equation}
  \beff(0) \;=\; \geff(0),
  \label{eq:ic_universal_threshold}
\end{equation}
to which it reduces whenever the clearance rate is prevalence-independent.

At the disease-free state, $\avg{II}=O(i^{2})$---co-infection requires two
independent transmission events---so $p\to1$ as $i\to0$. Formally, setting $i=0$
in Eq.~\eqref{eq:ic-pdot} gives
$(1-p)\bigl[\beff^{\mathrm{inter}}(0)+\gamma_2(1+p)-\gamma p\bigr]=0$, whose relevant root
in $[0,1]$ is
\begin{equation}
  p^{*}(0)=1 ,
\end{equation}
\emph{independently of $\gamma_2$}. Hence $\beta_{\mathrm{eff}}(0)=\tau$ and
$\gamma_{\mathrm{eff}}(0)=\gamma$: at vanishing prevalence there is no co-infection,
so neither the transmission enhancement $\alpha$ nor the co-infected clearance rate
$\gamma_2$ is felt, and each disease invades as an ordinary \SIS\ process. The
Jacobian of Eqs.~\eqref{eq:ic-idot}--\eqref{eq:ic-pdot} at $(i,p)=(0,1)$ is lower
triangular with invasion eigenvalue
$\lambda_1=\beta_{\mathrm{eff}}^{\mathrm{inter}}(0)-\gamma_{\mathrm{eff}}^{\mathrm{inter}}(0)=\tau-\gamma$, so
by Eq.~\eqref{eq:ic_universal_threshold}
\begin{equation}
    \tau_c=\gamma\;
  \label{eq:ic_threshold}
\end{equation}
independently of \emph{both} $\alpha$ and $\gamma_2$.

\paragraph{Step 3 — 1D reduction.}
Imposing the quasi-static condition $\dot p=0$ in Eq.~\eqref{eq:ic-pdot} and
using Eq.~\eqref{eq:ic_beff} gives a quadratic $A(i)\,p^{2}+B(i)\,p+C(i)=0$ with
\begin{equation}
  A=\tau(\alpha-1)(1+i)+(\gamma-\gamma_2),\qquad
  B=-\bigl[\gamma+\tau(2\alpha-1+i)\bigr],\qquad
  C=\gamma_2+\tau\alpha(1-i).
  \label{eq:ic_ABC}
\end{equation}
The asymmetry $\gamma-\gamma_2$ enters only the leading coefficient, and the
coefficients reduce to the symmetric ones when $\gamma_2=\gamma$. One structural
property survives untouched: $A(i)+B(i)+C(i)=-2\tau i$ vanishes at $i=0$, so the
physical root passes through $p^{*}(0)=1$ for every $\gamma_2$, consistent with
Step~2. Provided $A>0$---which holds for $\alpha>1$ and $\gamma_2\le\gamma$, and more
generally whenever co-infected clearance is not so fast as to overwhelm the
enhancement---the two roots straddle unity and the physical branch is
\begin{equation}
  p^{*}(i)=\frac{-B-\sqrt{\,B^{2}-4AC\,}}{2A}\in[0,1],
  \label{eq:ic_pstar}
\end{equation}
which closes the one-dimensional equation
\begin{equation}
  \dot i \;=\; i\left[\beta_{\mathrm{eff}}^{\mathrm{inter}}(i)-\gamma_{\mathrm{eff}}^{\mathrm{inter}}(i)\right]
\end{equation}
with
\begin{equation}
  \beta_{\mathrm{eff}}^{\mathrm{inter}}(i)=\tau\bigl[\alpha-(\alpha-1)p^{*}(i)\bigr](1-i),
  \quad
  \gamma_{\mathrm{eff}}^{\mathrm{inter}}(i)=\gamma_2+(\gamma-\gamma_2)\,p^{*}(i).
  \label{eq:ic_1dEQ}
\end{equation}
Evaluating $\beta_{\mathrm{eff}}^{\mathrm{inter}}(0)-\gamma_{\mathrm{eff}}^{\mathrm{inter}}(0)=0$ returns
$\tau_c=\gamma$.

\paragraph{Step 4 — Bifurcation.}
We apply Proposition~\ref{prop:transcrit} with $x=i$, $\lambda=\tau$ and
\begin{equation*}
  g(i,\tau)=\beta_{\mathrm{eff}}^{\mathrm{inter}}(i)-\gamma_{\mathrm{eff}}^{\mathrm{inter}}(i),
\end{equation*}
where $p^{*}(i,\tau)$ is the root of Eq.~\eqref{eq:ic_pstar}. The bifurcation sits at
$\tau_c=\gamma$ with $p^{*}(0)=1$. Implicit differentiation
of Eq.~\eqref{eq:ic_ABC} at $(i,p)=(0,1)$ gives
\begin{equation}
  \partial_i p^{*}(0,\tau)=\frac{2\tau}{\gamma-2\gamma_2-\tau},
  \qquad
  \partial_\tau p^{*}(0,\tau)=0,
  \label{eq:ic_pderivs}
\end{equation}
so that at threshold
\begin{equation}
  \partial_i p^{*}(0,\gamma)=-\frac{\gamma}{\gamma_2},
\end{equation}
which returns the symmetric value $-1$ when $\gamma_2=\gamma$. Hence
\begin{align}
  \partial_i g(0,\tau_c)
   &= \underbrace{-\tau(\alpha-1)\,\partial_i p^{*}(0,\gamma)-\beta_{\mathrm{eff}}(0)}
      _{\textstyle \beta_{\mathrm{eff}}'(0)=\gamma^{2}(\alpha-1)/\gamma_2 - \gamma}
      \;-\;\underbrace{(\gamma-\gamma_2)\,\partial_i p^{*}(0,\gamma)}
      _{\textstyle \gamma_{\mathrm{eff}}'(0)=-\gamma(\gamma-\gamma_2)/\gamma_2}
      \;=\;\frac{\gamma}{\gamma_2}\bigl(\alpha\gamma-2\gamma_2\bigr),
      \label{eq:ic_dxg}\\[4pt]
  \partial_\tau g(0,\tau_c)
   &= \bigl[\alpha-(\alpha-1)p^{*}(0)\bigr]
      -\tau(\alpha-1)\,\partial_\tau p^{*}(0,\gamma) \;=\; 1 .
\end{align}
By Proposition~\ref{prop:transcrit},
\begin{equation}
  \frac{\partial_i g(0,\tau_c)}{\partial_\tau g(0,\tau_c)}
  =\frac{\gamma(\alpha\gamma-2\gamma_2)}{\gamma_2},
  \qquad
  \Lambda'(0)=-\frac{\partial_i g}{\partial_\tau g}
  =\frac{\gamma}{\gamma_2}\bigl(2\gamma_2-\alpha\gamma\bigr),
\end{equation}
so the transition is forward (continuous) for $\alpha<\alpha^{*}$ and backward
(discontinuous, with bistability) for $\alpha>\alpha^{*}$, with boundary
\begin{equation}
  \;\alpha^{*}=\frac{2\gamma_2}{\gamma}\;
  \label{eq:ic_alphastar}
\end{equation}
which is $\alpha^{*}=2$ in the symmetric case. A
prolonged co-infection period ($\gamma_2<\gamma$) lowers the transmission
enhancement required for a discontinuous transition, and once
$\gamma_2<\gamma/2$ the boundary falls below $\alpha^{*}=1$. Bistability is then
possible even when co-infection \emph{reduces} transmission ($\alpha<1$), driven
purely by the longer time co-infected individuals spend infectious. This regime is, to our knowledge, so far ignored in the literature on interacting contagions.

\subsubsection{Pairwise SIR}
\label{sec:sir}

Every model considered so far collapses
to a single equation because knowing the prevalence $i$ specifies the susceptible population $s=1-i$.
However, \SIR\ dynamics does not as it intrinsically contains two degrees of freedom and one must specify both $i$ and $s$. 
This is simply a statement about which variables are actually fast: susceptible depletion proceeds on the \emph{same} timescale as the prevalence itself, so $s$ is a second \emph{slow} variable. 
The reduction therefore terminates at two dimensions.

\paragraph{Step 0 — Full system.}
The full pairwise SIR equations are
\begin{align}
\dot{\langle S \rangle}  &= -\tau n \langle SI \rangle, \label{eq:PW_SIR_nReg_full_S}\\
\dot{\langle I \rangle}  &=  \tau n \langle SI \rangle - \gamma \langle I \rangle, \label{eq:PW_SIR_nReg_full_I}\\
\dot{\langle R \rangle}  &=  \gamma \langle I \rangle, \label{eq:PW_SIR_nReg_full_R}\\[6pt]
\dot{\langle SI \rangle} &=  \tau(n-1)\langle SSI \rangle
- \tau(n-1)\langle ISI \rangle
- \tau \langle SI \rangle
- \gamma \langle SI \rangle, \label{eq:PW_SIR_nReg_full_SI}\\
\dot{\langle SS \rangle} &= -2\tau(n-1)\langle SSI \rangle
 \label{eq:PW_SIR_nReg_full_SS},
\end{align}
with closure relations
\begin{equation}
\langle SSI \rangle = \frac{\langle SS \rangle\langle SI \rangle}{\langle S \rangle} , \quad \langle ISI \rangle = \frac{\langle SI \rangle^2}{\langle S \rangle} \label{eq:PW_SIR_nReg_full_closures}
\end{equation}

\paragraph{Step 1 — Minimum system.}
Conservation laws are less powerful here because of the enlarged state space. 
The full five-equation system, Eqs.~\eqref{eq:PW_SIR_nReg_full_S}--\eqref{eq:PW_SIR_nReg_full_SS}, admits a modest reduction to four equations~\cite{Kiss2017},
\begin{align}
\dot{\langle S \rangle}  &= -\tau n \langle SI \rangle, \label{eq:PW_SIR_min_nReg_full_S}\\
\dot{\langle I \rangle}  &=  \tau n \langle SI \rangle - \gamma \langle I \rangle, \label{eq:PW_SIR_min_nReg_full_I}\\
\dot{\langle SI \rangle} &=  \tau\clSSI - \tau\clISI - \tau \langle SI \rangle - \gamma \langle SI \rangle,
\label{eq:PW_SIR_min_nReg_full_SI}\\
\dot{\langle SS \rangle} &= -2\tau \clSSI \label{eq:PW_SIR_min_nReg_full_SS}.
\end{align}

\paragraph{Step 2 — The $(i,s,\sigma,p)$ system and the epidemic threshold.}
Setting $i=\langle I\rangle$, $s=\langle S\rangle$, $\sigma=\langle SS\rangle$ and
$p=\langle SI\rangle/\langle I\rangle$, the system becomes
\begin{align}
\dot{i} &= i\left(\tau n p - \gamma\right), \label{eq:pw_SIR_ip_i}\\
\dot{s} &= -\tau n p\, i, \label{eq:pw_SIR_ip_s}\\
\dot{\sigma} &= -2\tau(n-1)\frac{\sigma p\, i}{s}, \label{eq:pw_SIR_ip_sigma}\\
\dot{p} &= \tau(n-1)\frac{\sigma}{s}\,p - \tau(n-1)\frac{p^{2}i}{s} - \tau p - \tau n p^{2}.
\label{eq:pw_SIR_ip_p}
\end{align}
At the disease-free state $(i,s,\sigma)=(0,1,1)$, Eq.~\eqref{eq:pw_SIR_ip_p} reduces to
$\dot p=\tau(n-2)p-\tau n p^{2}$, whose non-trivial root is
\begin{equation}
  p^{*}(0)=\frac{n-2}{n},
  \label{eq:sir_pstar0}
\end{equation}
the familiar statement that an infected node has one infected neighbour---the one that
infected it---and, on the $n$-regular graph, at most $n-1$ susceptible ones. The
characteristic equation for $p^{*}$ is \emph{linear}, and Eq.~\eqref{eq:pw_SIR_ip_i} gives
the threshold
\begin{equation}
  \tau_c (n-2)=\gamma\;.
  \label{eq:sir_threshold}
\end{equation}

\paragraph{Step 3 — An exact invariant eliminates $\sigma$.}
Dividing Eq.~\eqref{eq:pw_SIR_ip_sigma} by Eq.~\eqref{eq:pw_SIR_ip_s} removes $\tau$, $p$ and
$i$ altogether,
\begin{equation}
  \frac{d\sigma}{ds}=\frac{2(n-1)}{n}\,\frac{\sigma}{s},
\end{equation}
which integrates to the exact invariant
\begin{equation}
  \sigma = C\,s^{2(n-1)/n},
  \qquad\text{hence}\qquad
  \frac{\sigma}{s}=C\,s^{(n-2)/n},
  \label{eq:sir_invariant}
\end{equation}
with $C$ fixed by the initial data ($C=1$ for $s_0\to1$). This step involves \emph{no
approximation}. 
The system is now three-dimensional in $(i,s,p)$.

\paragraph{Step 4 — Constraining $p$: the 2D reduction.}
Only now do we apply the recipe. Setting $\dot p=0$ in Eq.~\eqref{eq:pw_SIR_ip_p} and
discarding the trivial root $p=0$ leaves a \emph{linear} equation for $p$, with root
\begin{equation}
  p^{*}(i,s)=\frac{(n-1)\dfrac{\sigma}{s}-1}{\,n+(n-1)\dfrac{i}{s}\,}
  \;=\;\frac{(n-1)\,C\,s^{(n-2)/n}-1}{\,n+(n-1)\,i/s\,},
  \label{eq:sir_pstar}
\end{equation}
which reduces to Eq.~\eqref{eq:sir_pstar0} at $(i,s)=(0,1)$. Substituting back closes the
system in the two \emph{slow} variables,
\begin{equation}
  \dot i = i\bigl[\beff^{\SIR}(i,s)-\gamma\bigr],\qquad
  \dot s = -\beff^{\SIR}(i,s)\,i,\qquad
  \beff^{\SIR}(i,s)=\tau n\,p^{*}(i,s).
  \label{eq:sir_2D}
\end{equation}
The effective force of infection now depends on the susceptible density as well as the
prevalence---the natural generalisation of $\beff(i)$ once a second slow variable is
present.

\begin{figure}[t]
\centering
\includegraphics[width=\textwidth]{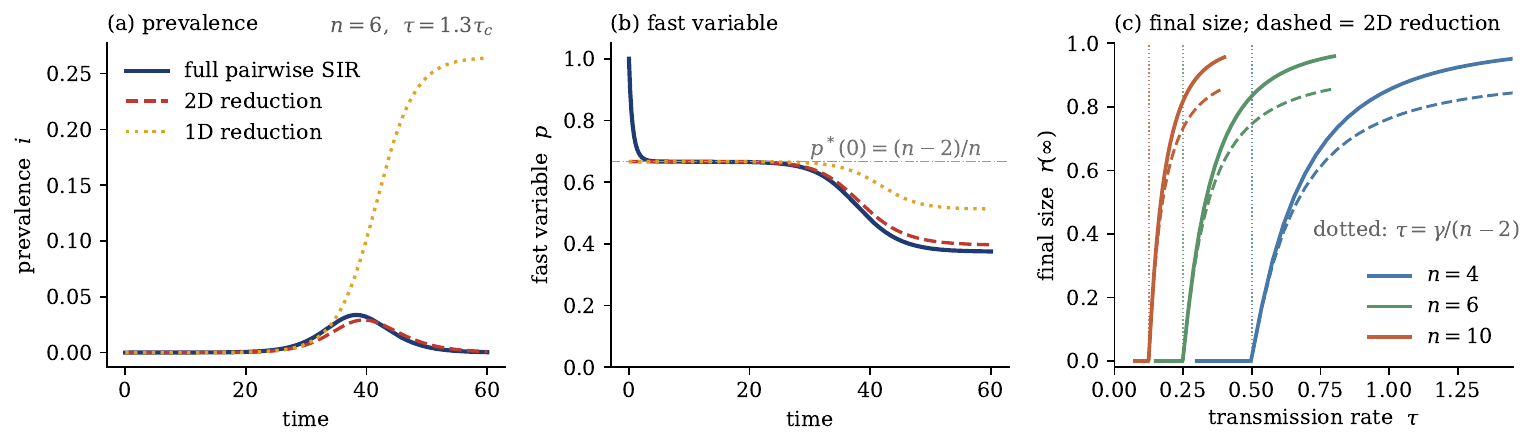}
\caption{\textbf{Pairwise \SIR: the reduction stops at two dimensions.}
(a) Prevalence $i(t)$ and (b) fast variable $p(t)$ for $n=6$, $\gamma=1$,
$\tau=2\tau_c$ with $\tau_c=\gamma/(n-2)$, seeded at $i_0=10^{-6}$.
The full pairwise \SIR\ model (solid) is compared with a 2D
reduction (dashed), and with a 1D
reduction assuming $s = 1$ (dotted). The grey line in (b) marks the quasi-static $p^{*}(0)=(n-2)/n$.
(c) Final size $r(\infty)$ against $\tau$ for $n\in\{4,6,10\}$: full model (solid) and
2D reduction (dashed), with the predicted thresholds $\tau=\gamma/(n-2)$ (dotted
verticals). The threshold is captured exactly; the final size to within $8$--$11\%$.}
\label{fig:sir_reduction}
\end{figure}

It is tempting to push on to a single equation by writing $s\to1-i$ and
$\sigma/s\to1$ in Eq.~\eqref{eq:sir_pstar}, which yields the closed form
\begin{equation}
  p(i)=\frac{(n-2)(1-i)}{n-i},
  \qquad
  \beff^{\SIR}(i)=\frac{\tau n (n-2)(1-i)}{n-i}.
  \label{eq:sir_pofI}
\end{equation}
This is a significant approximation, only valid around the critical point where $s$ stays close to 1. 
The threshold~\eqref{eq:sir_threshold}
is therefore recovered correctly, but agreement at the threshold certifies nothing about the trajectory (see Fig.~\ref{fig:sir_reduction}).

\subsection{Fast-variable reduction from simulations\label{sec:simulations}}
Here, we investigate how the fast-variable reduction can work in conjunction with data from simulations. We ask two questions. (1) To what extent do the fast-variable reduction and the fast-variable dynamics fit the behavior observed in stochastic simulations of the model? And (2) can we measure and fit the fast-variable from simulations to study the reduced 1D model?

All simulations are performed using a next-reaction method for Gillespie-style continuous-time simulation. 
For the all models, we generated five networks of size 10,000 that are: degree-regular for the \SIS, \SIR, and higher-order model, Poisson random networks for the adaptive network model, and fully connected for the model of interacting contagions.
We ran 20 independent simulations per network, for a total of 100 runs per model. 
All runs that go extinct before causing 100 infections are discarded as stochastic extinctions. 
To discard transient stochasticity, we align the remaining time series by calling $t=0$ the moment a given run reaches a threshold prevalence (5\% for the higher-order model, 1\% for all others).
In the noisy data coming from these simulations, we fit the quasi-static fast variable $p(i)$ using a quadratic function of the slow variable $i$ (and of both $i$ and $s$ for the \SIR~model: $p(i,s) = a_0 + a_1 i + a_2 s + a_3 i^2 + a_4 is + a_5 s^2$). With the result of this fit, we can then run our 1D model, tracking both prevalence $i$ and the the fast variable $p$ over time.

\begin{figure}
    \centering
    \includegraphics[width=\linewidth]{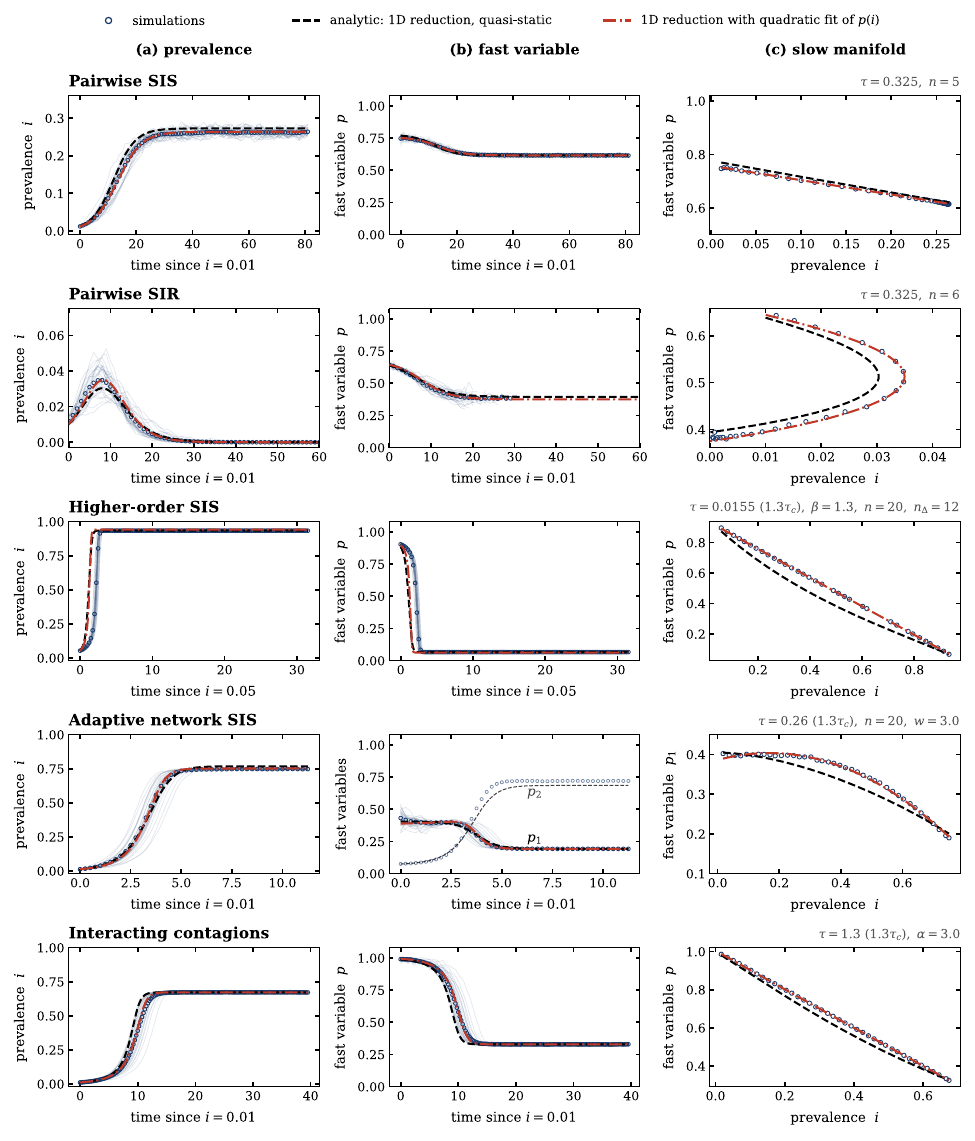}
    \caption{\textbf{Simulation-based quasi-static reduction for all models}. Panel (a) shows the simulations used as input (thin lines) and their average (markers), as well as the analytical 1D model (black) and the fitted 1D model (red). Panel (b) shows the behavior of the fast variable over time in simulations, in the analytical 1D model, and in the fitted 1D model. Panel (c) shows the core of this procedure: The quadratic fit of the fast variable $p$ as a function of prevalence $i$ and the corresponding quasi-static solution.}
    \label{fig:simulations}
\end{figure}

Figure \ref{fig:simulations} shows the results of this procedure. The fitted model, shown in red in all panels, agrees well with the simulations. We used denser networks for the adaptive network model as this is the regime where the mathematical description is meant to match simulations. The discrepancy between the model and simulations of the higher-order network model is also a known issue \cite{Malizia2025}, and the 1D reduction therefore performs as well as expected.

\end{document}